\documentclass{jpp}
\usepackage{graphicx}
\usepackage[utf8]{inputenc}
\usepackage[T1]{fontenc}
\usepackage{amsmath}

\usepackage{subcaption}
\usepackage{amssymb}
\usepackage{graphicx}
\usepackage{ulem}
\usepackage{esint}
\usepackage{float}
\usepackage{siunitx,physics}
\usepackage{flafter}

\shorttitle{W7-X microinstabilities and turbulence close to the stability threshold}
\shortauthor{L. Podavini et al.}

\title{Ion temperature and density gradient driven instabilities and turbulence in Wendelstein 7-X close to the stability threshold}

\author{L. Podavini\aff{1}
    \corresp{\email{linda.podavini@ipp.mpg.de}}, A. Zocco\aff{1}, J. M. Garc\'ia-Rega\~na\aff{2}, M. Barnes\aff{3}, F. I. Parra\aff{4}, A. Mishchenko\aff{1} \and P. Helander\aff{1}}

\affiliation{\aff{1}Max-Planck-Institut f\"ur Plasmaphysik, Wendelsteinstra{\ss}e 1, D-17491 Greifswald, Germany
\aff{2}Laboratorio Nacional de Fusi\'on CIEMAT, 28040 Madrid, Spain
\aff{3}Rudolf Peierls Centre for Theoretical Physics, University
of Oxford, Oxford OX1 3NP, United Kingdom
\aff{4}Princeton Plasma Physics Laboratory, 100 Stellarator Road, Princeton, USA}

\begin{document}

\maketitle

\begin{abstract}
Electrostatic gyrokinetic instabilities and turbulence in the Wendelstein 7-X stellarator are studied. Particular attention is paid to the ion-temperature-gradient (ITG) instability and its character close to marginal stability [Floquet-type turbulence \citep{PhysRevE.106.L013202} with no electron temperature gradient]. The flux-tube version of the $\delta f$ code \texttt{stella} \citep{barnes2019stella} is used to run linear and nonlinear gyrokinetic simulations with kinetic electrons. The nature of the dominant instability depends on the wavelength perpendicular to the magnetic field, and the results are conveniently displayed in stability diagrams that take this dependence into account. This approach highlights the presence of universal instabilities, which are less unstable but have longer wavelengths than other modes. A quasi-linear estimate of the heat flux suggests they are relevant for transport. Close to the stability threshold, the linear eigenmodes and turbulence form highly extended structures along the computational domain if the magnetic shear is small. Numerical experiments and diagnostics are undertaken to assess the resulting radial localisation of the turbulence, which affects the interaction of the latter with zonal flows. Increasing the amplitude of the magnetic shear (e.g. through current drive) has a stabilising effect on the turbulence and thus reduces the nonlinear energy transport.
\end{abstract}

\section{Introduction}
Stellarators and tokamaks are promising devices for achieving magnetically confined thermonuclear fusion. Poorly designed stellarators suffer from trapped-particle losses. As a result, they feature high levels of collisional (neoclassical) transport, significant power losses and poor confinement. These problems can be mitigated by optimising the geometry of the magnetic field \citep{nuhrenberg1986stable, boozer1995quasi}. A promising line of optimisation aims at achieving quasi-isodynamic magnetic fields \citep{nuhrenberg2010development}. Stellarators with this type of magnetic geometry are characterised by particle orbits with very small radial drifts, and thus low neoclassical transport.

Wendelstein 7-X (W7-X) is the first large, superconducting, optimised stellarator. The first experimental campaign proved the success of magnetic field optimisation. Indeed, W7-X shows a reduction of neoclassical transport effects and features among the highest energy confinement times for stellarators \citep{dinklage2018magnetic, bozhenkov2020high, Naturecraig}. However, power balance analyses highlight that the calculated remaining neoclassical transport is not sufficient to explain the observed energy losses. W7-X transport is found to be mainly turbulent and is thought to be dominated by ion-scale turbulence caused by microinstabilities driven by temperature and density gradients in the plasma \citep{bozhenkov2020high}. 

As in tokamaks, the microinstability which is thought to be most harmful to W7-X confinement is the one driven by the ion temperature gradient (ITG) \citep{carralero2021experimental}. Joint experimental and theoretical studies have highlighted how effects like the exacerbation of the turbulent heat transport with increasing electron heating can be explained by the presence of ITG-driven turbulence \citep{beurskens2021ion}. Decreasing the ion-to-electron temperature ratio $\tau=T_i/T_e$ at constant ion temperature $T_i$ lowers the ITG destabilisation threshold \citep{zocco2018threshold} and increases the turbulent transport. A similar effect has been observed in tokamaks too \citep{romanelli1989,xu1991numerical, casati2008temperature}. For W7-X, the importance of this effect has been confirmed by gyrokinetic simulations \citep{zocco2018threshold,beurskens2021ion} performed with the code \texttt{GENE} \citep{jenko2000electron}, and its influence on the ion-to-electron thermal coupling at transport scales has been investigated within a multiscale approach, providing a physical interpretation for the relatively low ion temperatures at high electron heating power \citep{BanonNavarro_2023}. 

The impact of the temperature ratio, $\tau$, on ITG modes can be studied in some detail close to the linear stability threshold. In this little-explored regime, it has been observed that the usual toroidal, fluid-like, branch of the ITG mode is replaced by a different type of ITG instability, of the Floquet-type \citep{bhatt,candywaltzrosen,jackbryanfloq,zocco2018threshold}. In this regime, the instability is characterised by an extended structure along the magnetic field lines. For this reason, in order for these modes to be observed in simulations, the parallel integration domain needs to be extended enough. As a result, these modes were overlooked in virtually all flux-tube stellarator simulations prior to the work in \citep{zocco2018threshold}.

Despite being of crucial relevance for transport, ITG turbulence in stellarators has rarely been studied in detail near its marginal stability threshold because of the relatively high computational requirements necessary for obtaining well-converged results.  Nonlinear gyrokinetic simulations performed with the code \texttt{stella} \citep{barnes2019stella}, however, show how such modes also cause a finite level of turbulent heat transport, if enough care is taken in resolving their extended parallel structure \citep{PhysRevE.106.L013202}.  

In \citep{PhysRevE.106.L013202}, the effects of density gradients on the nearly marginally stable ITG mode were neglected since density profiles in W7-X are often rather flat. Pellet injection, however, can be used to achieve higher densities and density gradients in high-performance discharges \citep{bozhenkov2020high}. These discharges show higher core ion temperatures and reduced turbulent fluctuations \citep{carralero2021experimental}. This behaviour was captured also through linear gyrokinetic simulations \citep{Alcuson_2020}. Here, we investigate previously partially explored regimes, producing high-fidelity near-marginal results, shading light on the wavelength dependence of the W7-X stability diagram, with particular emphasis on transport. Our highly resolved studies show that ITG and universal modes \citep{Coppi_1977,smoliakovkishimoto,landreman2015,helander-plunk2015} can be somewhat less unstable than ion-driven trapped-electron modes, but generally have longer wavelengths. A quasi-linear estimate of the turbulent heat flux suggests they are important for transport although they were overlooked in previously published stability diagrams \citep{Alcuson_2020}. Ion scale dynamics can, however, be impacted by the presence of electron temperature gradients \footnote{We address the importance of electron temperature gradients at the ion scale in W7-X in a dedicated work \citep[][in prep.]{zocco2024gradte}.}.

We successively focus on pure ITG instabilities with ion temperature gradients, but neglect electron temperature gradients. We investigate the parallel structure of the mode along the magnetic field line, and perform convergence studies to identify the most suitable input parameters for nonlinear simulations. A discussion follows on the numerical limits that make near-marginal studies difficult and why some compromises have to be made.

A full spectral analysis of fluctuations in the relevant regimes is presented, emphasising the difference between fluid-like and Floquet-like turbulence, especially during the pre-saturation and saturation phases. Differences are striking, and the regime identification is unambiguous. The study of Fourier spectra of fluctuations suggests that Floquet-type turbulence is more radially localised, and we speculate this is the reason why zonal flows are not efficient in shearing and suppressing turbulent structures \citep{PhysRevE.106.L013202}. To test this hypothesis, a numerical experiment is performed, where the zonal-flow contribution is switched off and the saturated amplitude of fluctuations is studied. We thus extend the previous work of \citep{PhysRevE.106.L013202}.

Finally, we further investigate the effect of the magnetic field geometry on the ITG linear threshold and transport levels. For this purpose we distort the radial profile of the rotational transform $\iota$ in accordance with a prescribed analytical function that exhibits strong variations, and recalculate the magnetic equilibria. We consider, in particular, the effect of electron cyclotron current drive (ECCD) \citep{zanini2020eccd, zanini2021confinement, aleynikova2021model, Zocco_2021} and observe indications of a possible increase of the stability threshold when $\iota$ is modified. It is found that the presence of strong magnetic shear regions in distorted $\iota$ profiles mitigates ITG modes, linearly and nonlinearly. More importantly, extended modes are successfully suppressed in this equilibria, leading to an increased ITG linear critical gradient.

The structure of this article is as follows. In Section \ref{sec:setting} we present the equations and numerical framework, and in the following section near-marginal linear and quasi-linear stability diagrams are presented. In Section \ref{sec:linear} and \ref{sec:nonlinear}, the linear and nonlinear results for the transition from fluid-like to Floquet-type ITG instabilities and turbulence are presented. Finally, in Section \ref{sec:shear}, the stabilising effects of $\iota$-profile distortions are studied through linear and nonlinear simulations performed in a non-vacuum magnetic equilibrium. In Section \ref{sec:conclusions} we summarise our results and conclusions.
 
\section{Physical and numerical setting}
\label{sec:setting}

Plasma microinstabilities are described mathematically by the gyrokinetic equation, which is an evolution equation for the gyrocentre distribution function $\delta G_s$, where $\delta G_s$ is the fluctuating part of the first-order correction to the equilibrium distribution function $f_s=F_{0s}+\delta f_s\equiv F_{0s}(1-Z_se\varphi(\vb{r},t)/T_{0,s})+\delta G_s(\vb{R}_s,\mu,\mathcal{E},t)+\order{\epsilon^2}$. $\varphi$ is the electrostatic potential, $Z_s e$ the electric charge and $T_{0s}$ the equilibrium temperature for the species $s$. For an electrostatic plasma the gyrokinetic equation reads:

\begin{equation}
\begin{split}
    & \left(\pdv{t}+v_{\parallel s}\grad_{\parallel}+\vb{v}_{ds}\cdot\grad\right)\delta G_s=\frac{Z_s eF_{0s}}{T_{0s}}\pdv{t}\left\langle\varphi\right\rangle_{\vb{R}_s}\\
    & -\frac{c}{B}\vu{b}\cdot\grad\left\langle\varphi\right\rangle_{\vb{R}_s}\cross\grad{F_{0s}}-\frac{c}{B}\vu{b}\cdot\grad{\left\langle\varphi\right\rangle_{\vb{R}_s}}\cross\grad{\delta G_s}\; .
\end{split}
\label{eq:gk}
\end{equation}

Collisions are neglected. The equilibrium distribution function $F_{0s}$ is taken to be Maxwellian and $\delta f_s/F_{0s}\sim k_{\parallel}/k_{\perp}\sim\rho^*\equiv\epsilon\ll 1$, where $k_{\parallel}$ and $k_{\perp}$ are wave numbers along and across the equilibrium magnetic field $\vb{B}$. $\rho^*\equiv\rho_s/L$ is the expansion parameter with $\rho_s=v_{ths}/\Omega_s$ the Larmor radius, $v_{ths}=\sqrt{2T_s/m_s}$ the thermal velocity and $L$ the equilibrium space scale. $\delta G_s$ depends on the gyrocentre position $\vb{R}_s=\vb{r}+\vb{v}_{\perp}\cross\vu{b}/\Omega_s$, with $\vb{r}$ being the particle position, $\Omega_s=Z_s e B/(m_sc)$ the gyrofrequency and $\vu{b}=\vb{B}/B$. The velocity-space coordinates are $\mu_s=m_sv_{\perp}^2/(2B)$ and $\mathcal{E}=m_s v^2/2$. The parallel gyrocentre velocity can thus be written as $v_{\parallel s}=\pm\sqrt{2(\mathcal{E}-B\mu_s)/m_s}$, while $\vb{v}_{ds}$ is the gyrocentre drift velocity $\vb{v}_{ds}=(v_{\parallel}/\Omega_s) \grad \cross (v_{\parallel}\vu{b})$, where $\grad=\partial/\partial \vb{R}_s$. The gyroaveraged electrostatic potential $\left\langle\varphi\right\rangle_{\vb{R}_s}$ is Fourier transformed with respect to $\vb{R}_s$, $\left\langle\varphi\right\rangle_{\vb{R}_s}=\sum_{\vb{k}}\left\langle\varphi\right\rangle_{\vb{R}_s,\vb{k}}\exp(i\vb{k}\cdot\vb{R}_s)$, where the coefficients are related to the coefficients of the transformation with respect to $\vb{r}$ through the Bessel function $\left\langle\varphi\right\rangle_{\vb{R}_s,\vb{k}}=J_0(a_s)\varphi_{\vb{k}}$, $a^2_s=2B\mu k_{\perp}^2/\Omega_s^2$. The electrostatic potential $\varphi$ is determined by the quasineutrality condition

\begin{equation}
    \sum_s Z_se\int\dd[3]{\vb{v}}F_{0s}\frac{Z_se\varphi}{T_{0s}}=\sum_s Z_se\int\dd[3]{\vb{v}}\sum_{\vb{k}}\text{e}^{i\vb{k}\cdot\vb{r}}J_0(a_s)\delta G_s\; .
\end{equation}

We perform numerical studies treating both ions and electrons kinetically, but neglecting electromagnetic effects. The code used is particularly suited for this task, as it is based on a semi-implicit numerical scheme. Our principal goal is to study the behaviour of gyrokinetic turbulence close to marginality. Simulating these regimes is demanding. For weakly-driven instabilities, a saturated turbulent state is difficult to reach, which translates into longer and computationally more expensive simulations than for strongly driven turbulence. For these reasons, we choose the flux-tube version of the $\delta f$ gyrokinetic code \texttt{stella} \citep{barnes2019stella}. This code solves the Fourier-decomposed nonlinear gyrokinetic equation (Eq. \ref{eq:gk}), multiplied by the normalising factor $(a^2/\rho_{ref} v_{th,ref})\exp(-v^2/v_{ths}^2)/F_{0s}$, where $v_{th,ref}$ denotes the thermal velocity of the user specified reference particle species, having mass $m_{ref}$ and temperature $T_{ref}$. Furthermore, $v_{ths}$ is the thermal velocity of the particle species $s$, $\rho_{ref}=v_{th,ref}/\Omega_{ref}$ is the Larmor radius of the reference particle species, where $\Omega_{ref}=Z_{ref} e B_{ref}/(m_{ref} c)$ is the gyrofrequency, and $a$ and $B_{ref}$ are, respectively, the reference length and reference magnetic field, which depend on the choice of the magnetic geometry. 
 
The coordinates used by \texttt{stella} are $(x,y,z,v_{\parallel},\mu)$, where $(x,y)$ are, respectively, the radial and binormal coordinates in the plane perpendicular to $\vu{b}$. The magnetic field vector is expressed as $\vb{B}=\grad{\alpha}\times\grad{\psi}$, where $\alpha = \theta - \iota\zeta$ is a coordinate that labels field lines and $\psi$ flux surfaces. Here, $\iota$ is the rotational transform, and $\theta$ and $\zeta$ denote poloidal and toroidal PEST flux coordinates \citep{grimm1983ideal}, respectively. In terms of these coordinates, $x$ and $y$ are defined as
\begin{equation}
\begin{split}
    & x = r-r_0\\
    & y = r_0(\alpha-\alpha_0)\; ,
\end{split}
\end{equation}
where $r_0$ and $\alpha_0$ are input parameters that specify the radial position and the field line at the centre of the simulation domain. In particular, $r$ is defined as $r\equiv a\sqrt{\psi_t/\psi_{t,LCFS}}$, where $\psi_t$ is the toroidal magnetic flux and $\psi_{t,LCFS}$ its value at the last closed flux surface. The perpendicular wave number $\vb{k}_{\perp}$ can thus be decomposed as $\vb{k}_{\perp}=k_x\grad{
x}+k_y\grad{y}$, where $k_x$ is the radial wave number and $k_y$ the binormal wave number. Furthermore, 
$z$ is a coordinate measuring the location along the magnetic field and, for stellarator simulations, is equal to the toroidal angle $\zeta$ measured in radians. 

All the simulations presented in this paper are run in stellarator geometry. The equilibrium is passed to \texttt{stella} as an input file obtained through the equilibrium solver code \texttt{VMEC} \citep{hirshman1983steepest}. The simulations are performed either in the W7-X high-mirror (KJM) configuration \citep{geiger2014physics} or in the low-mirror (AIM) one. The main difference between these two configurations is their mirror ratio, i.e. the ratio between the maximum and minimum magnetic field strength on the magnetic axis. The high-mirror configuration is of particular interest for experiments as it is characterised by a small bootstrap current.

\section{Stability diagrams}
\label{sec:map}

\begin{figure}
    \centering
    \includegraphics[scale=0.45]{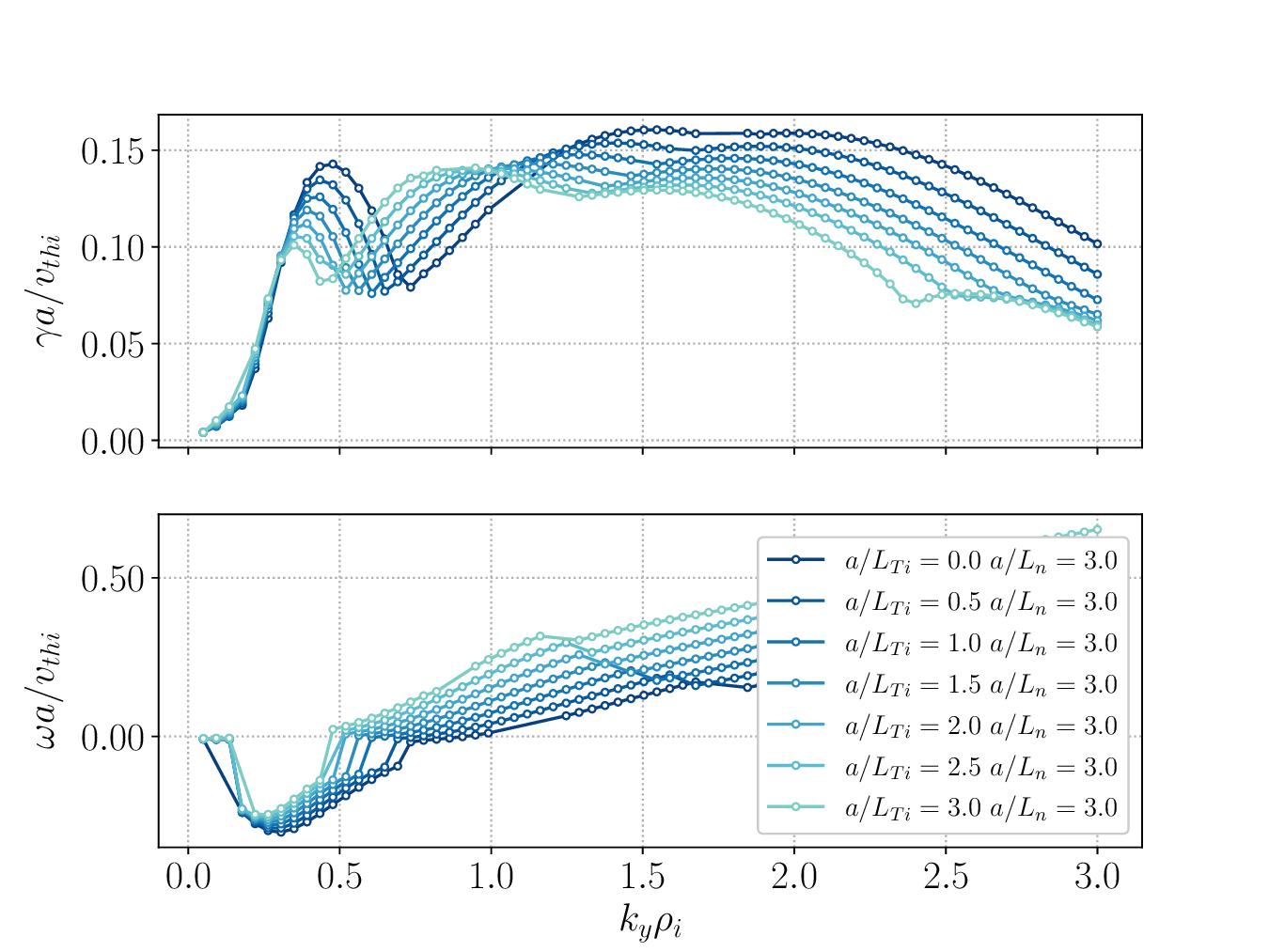}
    \caption{Normalised growth rate $\gamma a/v_{thi}$ and frequency $\omega a/v_{thi}$ spectra as a function of the normalised binormal wave number $k_y\rho_i$ for simulations with density gradient $a/L_n=3.0$ but different ion temperature gradients $a/L_{Ti}$.}
    \label{fig:fprim3}
\end{figure}

We begin by constructing a stability diagram for the various electrostatic microinstabilities that could harm plasma confinement, in the spirit of \citep{Alcuson_2020} but carefully monitoring the wave number dependence of the various modes. Selecting the most unstable eigenvalue of  microinstability spectra is not always the most conclusive procedure, since the relevance of an instability for transport depends on the scale at which the instability is active. We perform a scan over the normalised ion temperature gradient $a/L_{Ti}$ and normalised density gradient $a/L_{n}$, where $1/L_{Ti}=-T^{-1}_{0i}dT_{0i}/dx$ and $1/L_{n}=-n^{-1}_{0}dn_{0}/dx$ are the characteristic gradients length scales and $a$ is the average minor radius of the plasma. The scan is performed in the range $[0.0, 3.0]$ for both gradients, with a step size of 0.5. Since we are simulating two kinetic species, we expect to observe ITG modes,  density-driven trapped electron modes (TEMs) \citep{helander2015advances, PhysRevLett.108.245002}, ion-driven trapped electron modes (iTEMs) \citep{proll2015tem, plunk_connor_helander_2017}, and universal instabilities \citep{Coppi_1977,smoliakovkishimoto,landreman2015,helander-plunk2015}. We consider equal temperatures $\tau\equiv T_i/T_e=1$ and density gradients. We set the electron temperature gradient to zero. We acknowledge the importance that electron temperature gradients can have at the ion scale in predominantly electron heated plasmas, such as in W7-X \citep{wilms2023full}. However, in this work we decide to limit ourselves to at most density gradient driven electron modes and their effect on ion physics. The role of electron temperature gradients in W7-X will be addressed in dedicated studies \citep[][in prep.]{zocco2024gradte}.

\begin{figure}
    \centering
    \includegraphics[scale=0.57]{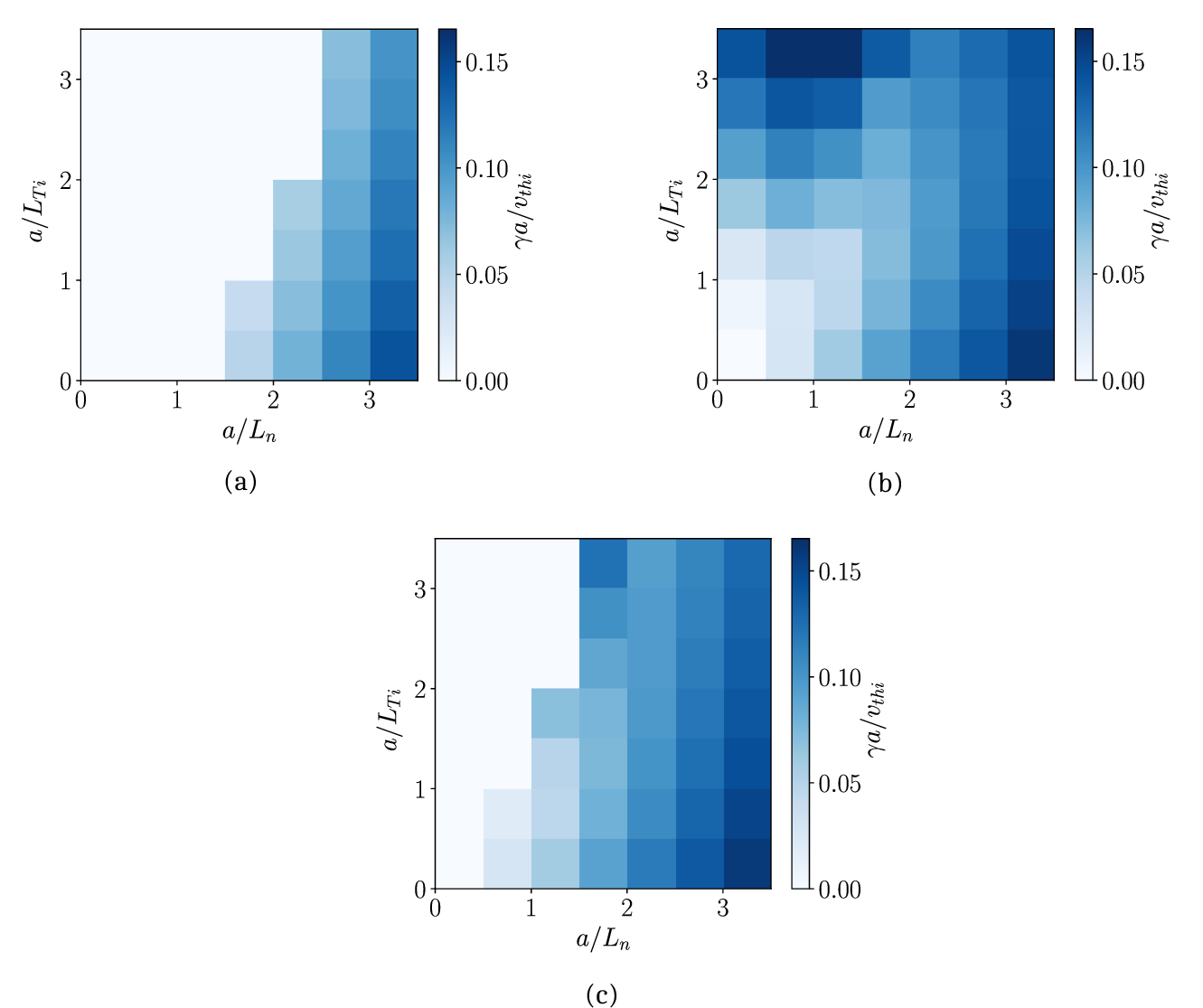}
    \caption{Stability diagrams showing the normalised growth rate $\gamma a/v_{thi}$ as a function of the ion temperature gradient $a/L_{Ti}$ and density gradient $a/L_n$. Each diagram shows instabilities having different normalised frequency $\omega a/v_{thi}$ signs and active at different $k_y\rho_i$: (a) $\omega a/v_{thi}<0$, $k_y\rho_i<1.0$, (b) $\omega a/v_{thi}>0$, $k_y\rho_i \lesssim 1.5$ and (c) $\omega a/v_{thi}>0$, $1.5 \lesssim k_y\rho_i \lesssim 2.5$.}
    \label{fig:all_maps}
\end{figure}

The radial location chosen is $r_0/a=0.7$, where electrostatic fluctuations have been detected in W7-X \citep{bahner2021phase}. The chosen equilibrium is a high-mirror (KJM) W7-X magnetic configuration. At this radial location, for the chosen equilibrium, the local value of the global magnetic shear is $\hat{s}=-(r/\iota) \dd \iota/\dd r=-0.1249$ and the safety factor $q=1/\iota=1.103$. The resolution of the simulations is set by the input parameters: $N_z\cross N_{k_y}\cross N_{k_x}\cross N_{v_{\parallel}}\cross N_{\mu}=384\cross 70\cross 1\cross 64\cross 24$. The maximum magnetic moment $\hat{\mu}_{max}$ is set by the maximum perpendicular velocity $\hat{v}_{\perp, max}=3$. The maximum parallel velocity is $|\hat{v}_{\parallel, max}|=3$, where $\hat{\mu}_s=\mu_s B_{ref}/(2T_s)$, $\hat{v}_{\perp s}=v_{\perp s}/v_{ths}$ and $\hat{v}_{\parallel s}=v_{\parallel s}/v_{ths}$. The number of binormal wave numbers corresponds to a scan in the range $k_y\rho_i\in [0.05,3.0]$, while the radial wave number $k_x\rho_i$ is set to zero. We choose to run the simulations in the flux-tube denoted by $\alpha_0=0$, centred around $(\theta, \zeta)=(0,0)$, at the bean-shaped cross section, which has been characterised extensively \citep{gonzalez2022electrostatic} and where turbulence is found to be significant \citep{helander2012stellarator}. The flux-tube length covers three full toroidal turns of the device for reasons to be discussed later. 

We construct three types of stability diagrams. This is because care must be taken to avoid overlooking instabilities with growth rates smaller than, but comparable to, the fastest growing one, especially at long wavelengths. Different instabilities are recognised by the sign of their normalised frequency, which may exhibit sudden jumps as the binormal wave number $k_y\rho_i$ is varied, since the code always selects the fastest growing instability. The three stability diagrams reflect the three main $k_y\rho_i$ ranges where instabilities are found, as displayed in Fig.~\ref{fig:fprim3}. At the longest wavelengths, $k_y\rho_i<1.0$, we find instabilities with negative frequency, which corresponds to rotation in the electron diamagnetic direction. At shorter wavelengths, we find two distinct instabilities having positive frequency, thus rotating in the ion diamagnetic direction. They are unstable for different $k_y\rho_i$: $k_y\rho_i\lesssim 1.5$ and $1.5\lesssim k_y\rho_i\lesssim 2.5$, respectively. We thus separate them into different diagrams, shown in Figs.~\ref{fig:all_maps}a-\ref{fig:all_maps}c, which are devised to show the normalized growth rate for the fastest growing mode in a selected $k_y\rho_i$ range. Each pixel corresponds to a single simulation having a specific value of the parameter $\{a/L_{Ti}, a/L_n\}$.

\begin{figure}
    \centering
    \includegraphics[scale=0.45]{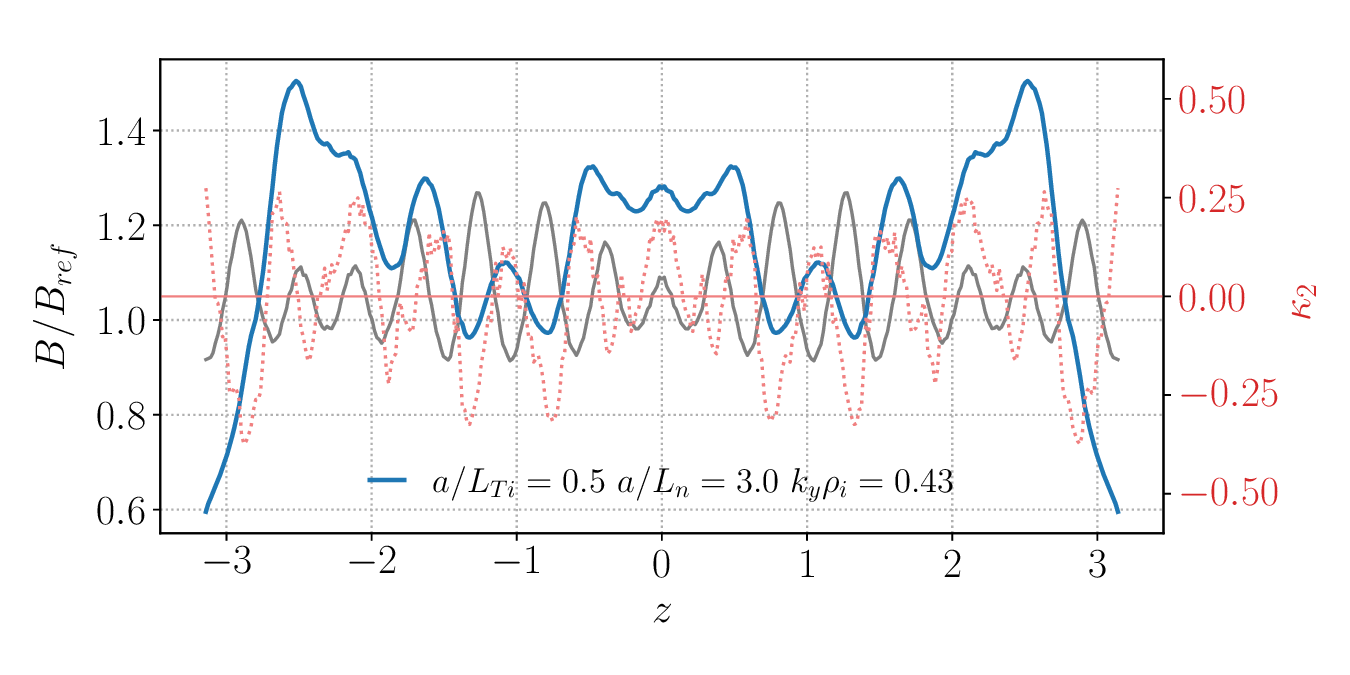}
    \caption{Electrostatic potential structure along the magnetic field line for $k_y\rho_i=0.43$, for the simulation with $\{a/L_{Ti}, a/L_n\}=\{0.5,3.0\}$ -- in blue -- compared with the normalised magnetic field strength $B/B_{ref}$ -- in grey -- and the magnetic drift $\kappa_2$ -- in dotted red.}
    \label{fig:tprim05_fprim3}
\end{figure}

The first diagram (Fig.~\ref{fig:all_maps}a) is populated by instabilities found at large scales and rotating in the electron diamagnetic direction. A critical value of the density gradient $a/L_n\approx 1.5$ is necessary to destabilise them, while the ion temperature gradient plays a stabilising role. In order to identify these instabilities, we study the parallel structure of the modulus of the normalised electrostatic potential $|\hat{\varphi}_{\vb{k}}|$, where $\hat{\varphi}_{\vb{k}}=(a/\rho_i)(e/T_i)\varphi_{\vb{k}}$. In particular, we compare it to two quantities useful for the instabilities identification: the value of the normalised magnetic field $B/B_{ref}$ and the normalised $\grad{B}$--drift in the $\grad{\alpha}$ direction $2a^2B_{ref}B^{-2}(\vu{b}\times\grad{B})\cdot\grad{\alpha}$, which we will denote here as $\kappa_2$. The former shows the presence of magnetic wells, while from the latter the regions of ‘‘bad curvature'' can be identified -- here denoted by positive values of the magnetic drift. 

\begin{figure}
    \centering
    \includegraphics[scale=0.45]{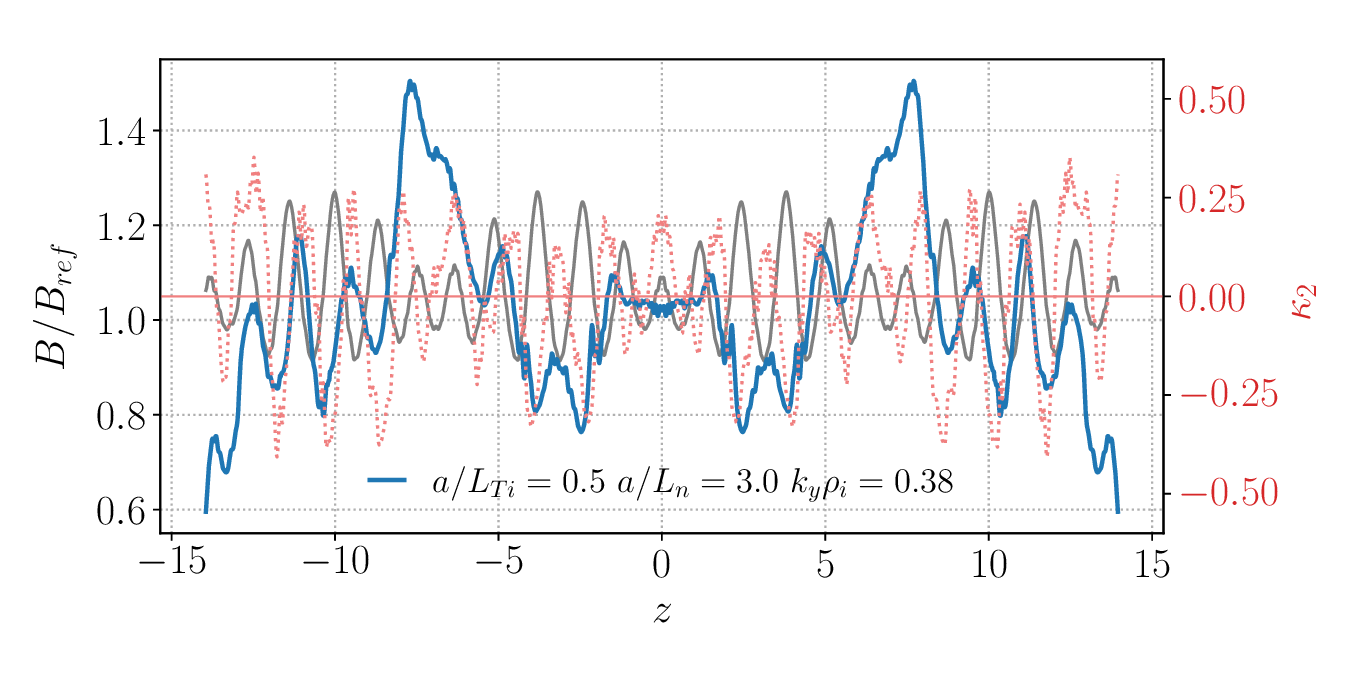}
    \caption{Electrostatic potential structure along the magnetic field line for $k_y\rho_i=0.38$, for the simulation with $\{a/L_{Ti}, a/L_n\}=\{0.5,3.0\}$ -- in blue -- compared with the normalised magnetic field strength $B/B_{ref}$ -- in grey -- and the magnetic drift $\kappa_2$ -- in dotted red. Result for a flux-tube that extends four times toroidally.}
    \label{fig:tprim05_fprim3_Nz1500}
\end{figure}

Fig.~\ref{fig:tprim05_fprim3} shows the electrostatic potential for one of the simulations composing the first diagram. We notice that the eigenfunction peaks in regions of bad curvature and is characterised by two symmetric maxima at finite $z$. The structure along $z$ is broad and decays slowly along the field line. A comparison with previous works suggests that we can identify them as universal instabilities \citep{Coppi_1977,smoliakovkishimoto,landreman2015,helander-plunk2015}. In contrast to earlier numerical simulations \citep{costello2023universal}, we find them also in the presence of a non-zero ion temperature gradient, which somewhat stabilises them but does not suppress them completely. We stress that the simulation of a long flux-tube was necessary to obtain this result. Unless multiple toroidal turns are included in the simulation domain, it is not possible to resolve the maxima at finite $z$. We show in Fig.~\ref{fig:tprim05_fprim3_Nz1500} that if an additional toroidal turn is simulated, other, less pronounced local maxima are found, but overall the eigenfunction decays along the field line. Three toroidal turns are thus enough to capture the main structure of the mode, and four are useful for obtaining a substantial decay of the eigenfuction towards the ends of the domain.

\begin{figure}
    \centering
    \includegraphics[scale=0.45]{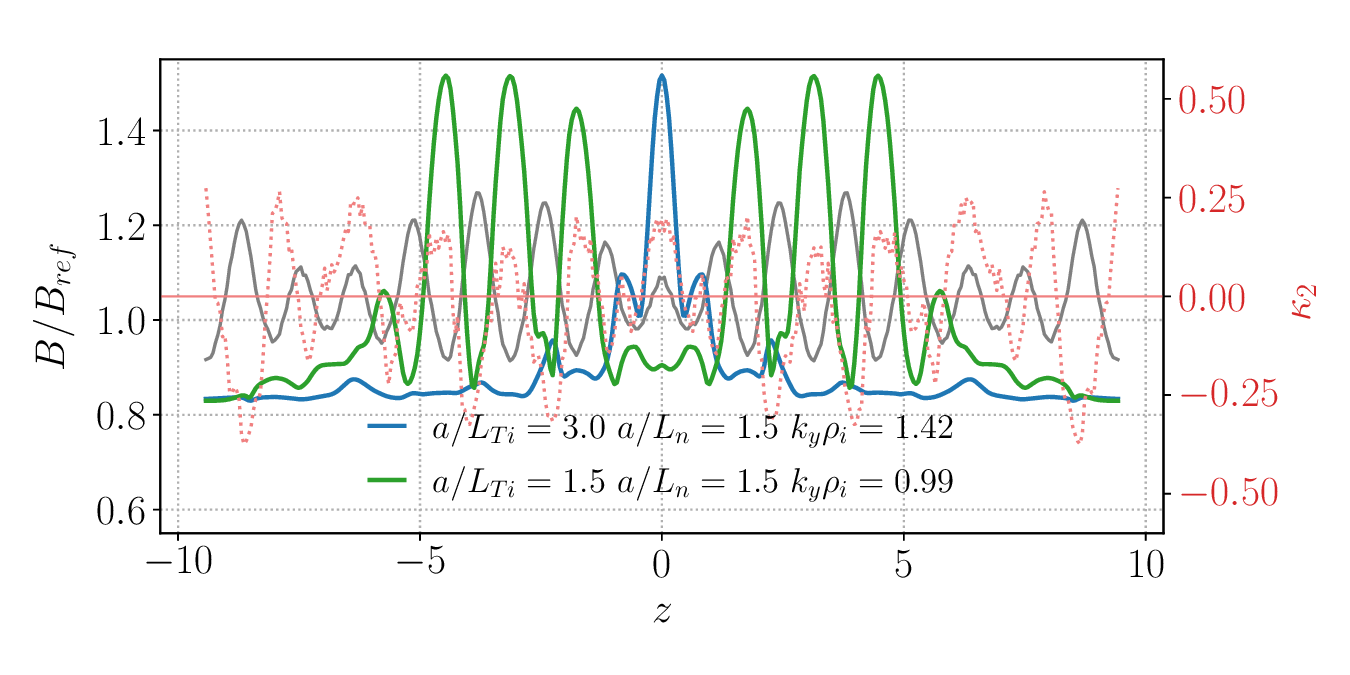}
    \caption{Electrostatic potential structure along the magnetic field line for $k_y\rho_i=0.99$ -- in green -- and $k_y\rho_i=1.42$ -- in blue -- for simulations with $a/L_{Ti}=\{1.5,3.0\}$ and $a/L_n=1.5$, compared with the normalised magnetic field strength $B/B_{ref}$ -- in grey -- and the magnetic drift $\kappa_2$ -- in dotted red.}
    \label{fig:tprim_fprim15}
\end{figure}

The second diagram (Fig.~\ref{fig:all_maps}b) is populated by instabilities with wave numbers up to  $k_y\rho_i \lesssim 1.5$ rotating in the ion direction. On the right side of the diagram, for $a/L_n \ge 1.5$, we mainly find iTEM instabilities, characterised by eigenfunctions that peak where magnetic wells and bad curvature regions overlap (see Fig.~\ref{fig:tprim_fprim15}). These instabilities are not very sensitive to variations in the ion temperature gradient as long as $a/L_n \ge 1.5$, but below this value the growth rate increases with increasing ion temperature gradient $a/L_{Ti}$. As we can observe in Fig.~\ref{fig:tprim_fprim15}, this can be explained by a transition from an iTEM to an ITG mode. The latter can be recognised by its strong peaking in the region of bad curvature at $z=0$. When the density gradient is further decreased, the ITG instability gradually comes to dominate for every value of $a/L_{Ti}$ and the growth rate becomes sensitive to the value of $a/L_{Ti}$, as expected.

\begin{figure}
    \centering
    \includegraphics[scale=0.45]{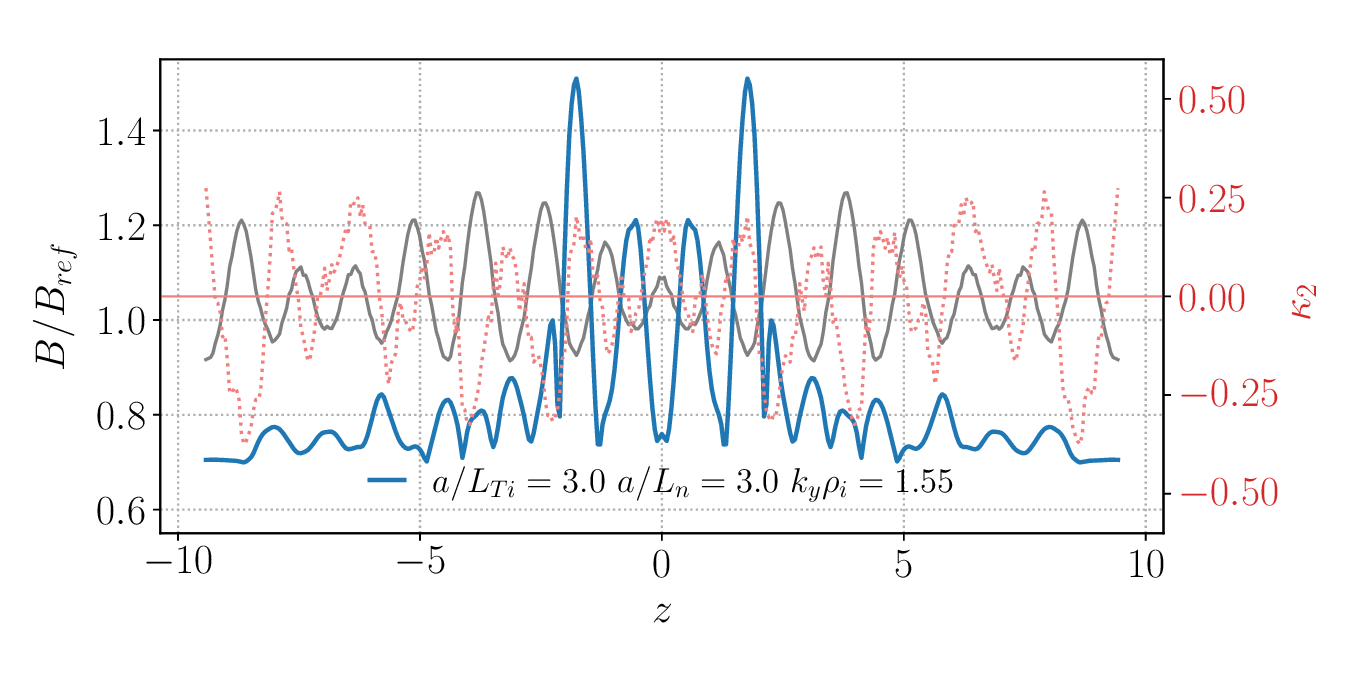}
    \caption{Electrostatic potential structure along the magnetic field line for $k_y\rho_i=1.55$, for the simulation with $\{a/L_{Ti}, a/L_n\}=\{3.0,3.0\}$ -- in blue -- compared with the normalised magnetic field strength $B/B_{ref}$ -- in grey -- and the magnetic drift $\kappa_2$ -- in dotted red.}
    \label{fig:tprim3_fprim3_map1}
\end{figure}

The third diagram (Fig.~\ref{fig:all_maps}c) describes instabilities that are found at smaller scales and rotate in the ion diamagnetic direction. In the right part of the diagram, the growth rate increases with the density gradient and slightly decreases with increasing ion temperature gradient. Fig.~\ref{fig:tprim3_fprim3_map1} shows an example of such an instability, which features local maxima corresponding to magnetic wells locations, and a local minimum in $z=0$. This type of mode does not possess any specific feature that could characterise it as ITG or iTEM, but rather a hybrid mode between the two. For $a/L_n<2$, however, the growth rate increases again with increased $a/L_{Ti}$. This suggests that a branch of the ITG mode is destabilised at smaller scales too.   

\begin{figure}
    \centering
    \includegraphics[scale=0.45]{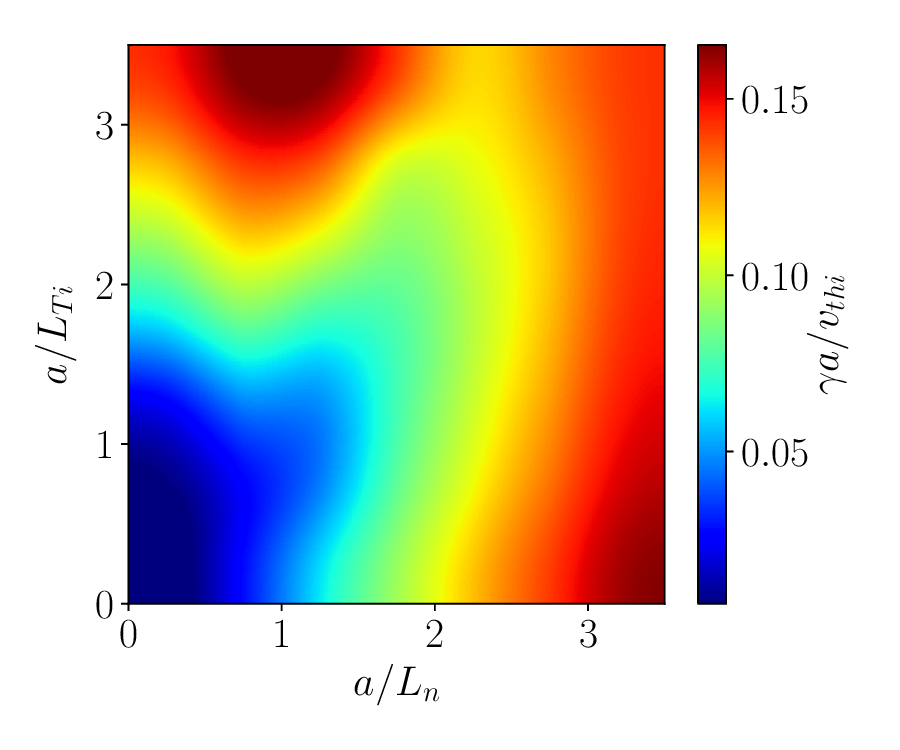}
    \caption{Stability diagram showing the normalised growth rate $\gamma a/v_{thi}$ as a function of the ion temperature gradient $a/L_{Ti}$ and density gradient $a/L_n$. Plotted here is the largest growth rate for each set of $\{a/L_{Ti}, a/L_n\}$, on which an interpolation has been performed.}
    \label{fig:general_map}
\end{figure}

\begin{figure}
    \centering
    \includegraphics[scale=0.57]{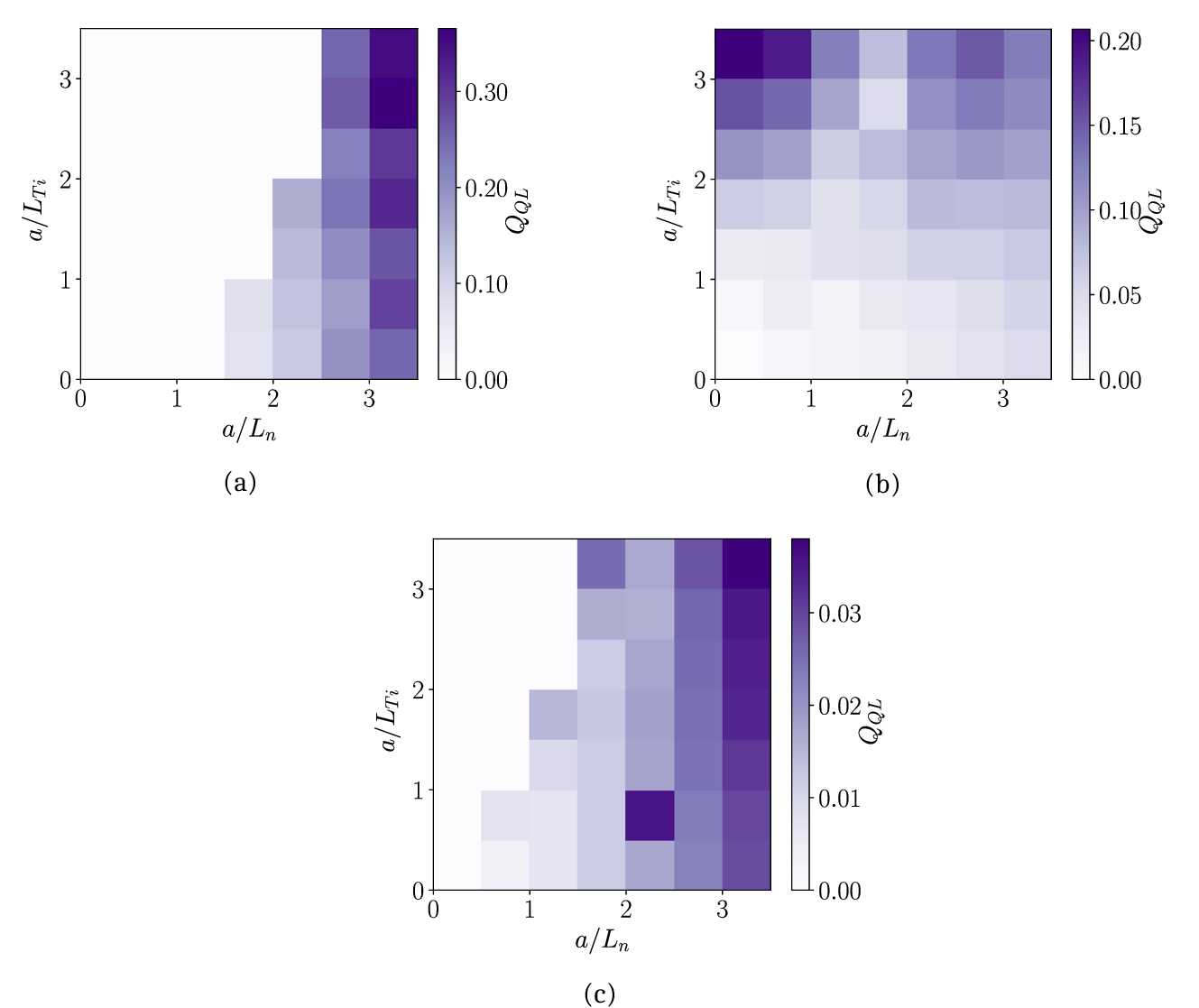}
    \caption{Stability diagrams showing the quasi-linear heat flux estimate for the instabilities reported in Figs.~\ref{fig:all_maps}a-\ref{fig:all_maps}c calculated as in Eq.~\ref{eq:ql}. Each diagram refers to instabilities having different normalised frequency $\omega a/v_{thi}$ signs and active at different $k_y\rho_i$: (a) $\omega a/v_{thi}<0$, $k_y\rho_i<1.0$, (b) $\omega a/v_{thi}>0$, $k_y\rho_i \lesssim 1.5$ and (c) $\omega a/v_{thi}>0$, $1.5 \lesssim k_y\rho_i \lesssim 2.5$.}
    \label{fig:all_maps_ql}
\end{figure}

We finally combine the three diagrams into one that only shows the growth rate for the fastest growing mode in the whole simulated $k_y\rho_i$-domain, as a function of $a/L_{Ti}$ and $a/L_n$ (Fig.~\ref{fig:general_map}). The comparison with the distinct diagrams shows how essential it is to separate different scales in order to understand which instabilities are active at which scales and for which gradients. The universal modes present in Fig.~\ref{fig:all_maps}a, for example, feature slightly smaller growth rates than other instabilities and are thus overlooked when only considering the fastest growing mode. But since they are active at larger scales, they might be more relevant for transport than modes of shorter wavelength.

In order to check our hypothesis, we perform a quasi-linear estimate for the heat flux \citep{mariani2018identifying}

\begin{equation}
    Q_{QL}=\frac{\hat{\gamma}}{\langle k_{\perp}^2\rangle}\; ,
\end{equation}
\label{eq:ql}

where $\hat{\gamma}=\gamma a/v_{thi}$ is taken from Figs.~\ref{fig:all_maps}a-\ref{fig:all_maps}c and

\begin{equation*}
\langle k_{\perp}^2\rangle=\frac{\int k_{\perp}^2|\varphi|^2\dd l/B}{\int|\varphi|^2\dd l/B}\; .
\end{equation*}

The result is reported in Figs.~\ref{fig:all_maps_ql}a-\ref{fig:all_maps_ql}c and is consistent with the hypothesis that universal modes can play a pivotal role. We observe that, while their growth rate is a decreasing function of $a/L_{Ti}$, $Q_{QL}$ is persistent for large $a/L_{Ti}$ (Fig.~\ref{fig:all_maps_ql}a). ITG modes, top left of Fig.~\ref{fig:all_maps_ql}b, are still relevant to transport, but not as much as universal instabilities. iTEMs, bottom right of Fig.~\ref{fig:all_maps_ql}b, on the other hand, do not seem to be as important. All remaining instabilities in Fig.~\ref{fig:all_maps_ql}c show a quasi-linear estimate more than ten times smaller than those already discussed in Figs.~\ref{fig:all_maps_ql}a and \ref{fig:all_maps_ql}b. 

A feature of Fig.~\ref{fig:general_map} is the presence of a ``stability valley'' at large gradients: as already noticed in \citep{helander2015advances} and more clearly emphasised in \citep{Alcuson_2020}, the growth rate does not increase with increasing gradients everywhere in the diagram. In particular, the addition of a non-zero density gradient to a plasma with a moderate ion temperature gradient reduces the growth rate. The term ``stability valley'' was coined by \citet{Alcuson_2020}, who calculated stability diagrams up to quite large gradients, which brings out this feature particularly clearly. For realistic values of the gradients, however, the ``valley'' is shallow and the growth rate only drops slightly. Perhaps a more important feature is that the fastest growing mode acquires a shorter wavelength, which reduces the turbulent transport \citep{helander2015advances,Xanthopoulos_2020}. Indeed, despite the mild decrease of the growth rate at experimentally relevant parameters, nonlinear simulations show a strong decrease of fluctuations and particle fluxes \citep{garcia2021turbulent2} along with heat fluxes \citep[][in prep.]{thienpondt2022turbulent, thienpondt2024turbulent}.

At large values of the density gradient, i.e., in the right of Fig.~\ref{fig:general_map}, there are rapidly growing universal and iTEM instabilities with long and short wavelengths, respectively. These instabilities were observed but not classified by  \citep{Alcuson_2020}. 

Measurements of W7-X temperature and density profiles suggest that normalised temperature gradients in most of the core do not exceed values of $a/L_{Ti}\approx 3.0$ and density gradients of $a/L_n\approx 1.5$. Such regimes are described by the left part of Fig.~\ref{fig:all_maps}b, which is dominated by ITG instabilities. For this reason, in the next sections, we will present an extensive study on the ITG, focusing in particular on its behaviour close to its linear stability threshold.

\section{Linear characterisation of the ITG mode near marginality}
\label{sec:linear}

In order to study the ITG mode, we run further linear, flux-tube, electrostatic simulations. We retain electron kinetic effects so as to have two kinetic species with equal temperatures. As seen in the stability diagram, the ITG mode is the dominant instability when the density gradient is small enough. Accordingly, we consider the limit of flat density profile ($a/L_n=0$) and also set $a/L_{Te}=0$, performing a scan in the normalised ion temperature gradient $a/L_{Ti}$ only, with values $a/L_{Ti}\in [0.5,3.0]$. Our aim is to ascertain whether ITG-driven modes, previously observed with adiabatic electrons \citep{zocco2018threshold}, persist when the electron dynamics is accounted for. We run our simulations in the same flux-tube used for the stability diagram,  $\alpha_0=0$, at the radial location $r_0/a=0.7$ and centered around $(\theta,\zeta)=(0,0)$ in the W7-X high-mirror magnetic configuration. Its extension is three toroidal turns. The resolution of the simulations is set by the input parameters: $N_z\cross N_{k_y}\cross N_{k_x}\cross N_{v_{\parallel}}\cross N_{\mu}=256\cross 62\cross 1\cross 64\cross 24$. The maximum magnetic moment $\hat{\mu}_{max}$ is set by the maximum normalised perpendicular velocity $\hat{v}_{\perp, max}=3$. The maximum normalised parallel velocity is $|\hat{v}_{\parallel, max}|=3$. The normalised binormal wave number is varied over the range $k_y\rho_i\in [0.05,2.0]$ whereas the normalised radial wave number $k_x\rho_i$ is set to zero. 

\begin{figure}
    \centering
    \includegraphics[scale=0.45]{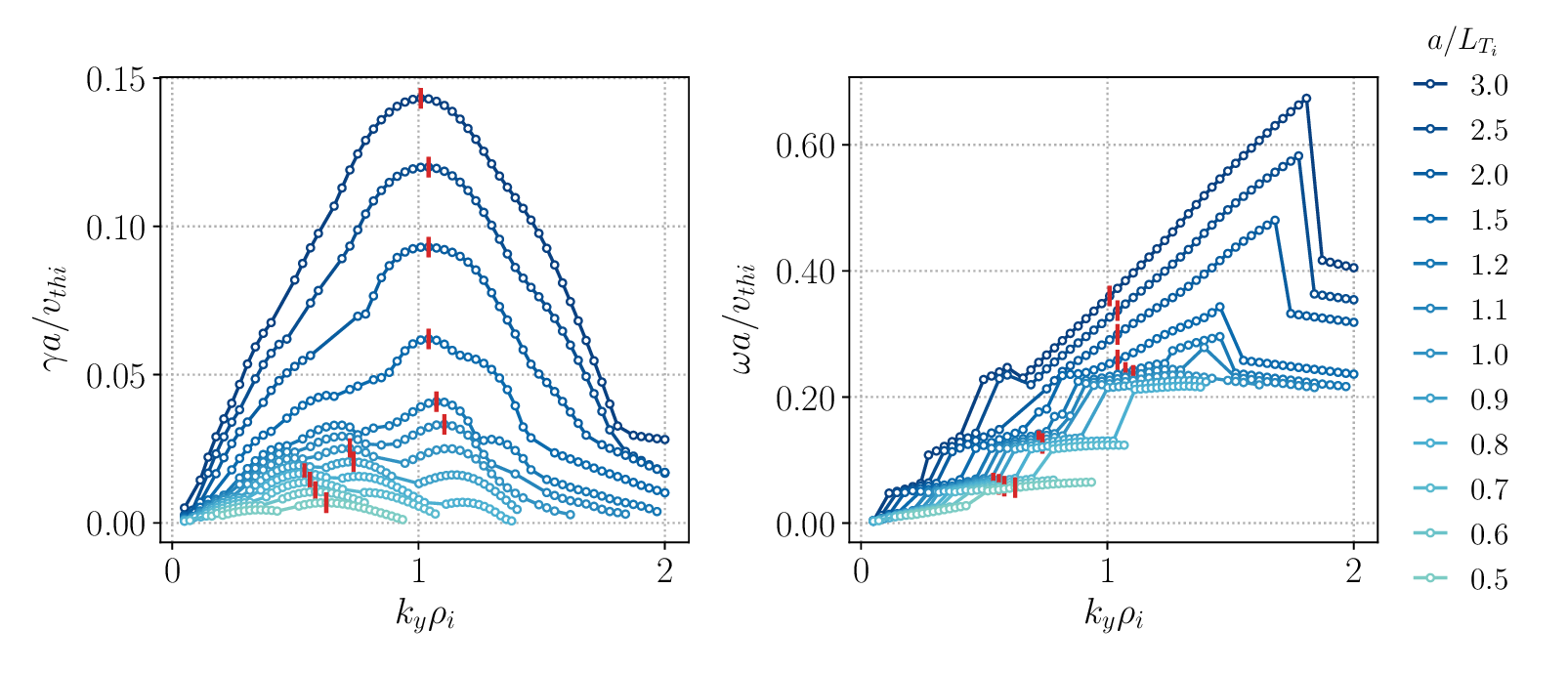}
    \caption{Spectra of the normalised linear growth rate $\gamma a/v_{thi}$ and frequency $\omega a/v_{thi}$ as a function of the normalised binormal wave number $k_y\rho_i$ for different temperature gradients. The growth rate and frequency for the fastest growing mode are highlighted.}
    \label{fig:spectra_new}
\end{figure}

\begin{figure}
    \centering
    \includegraphics[scale=0.45]{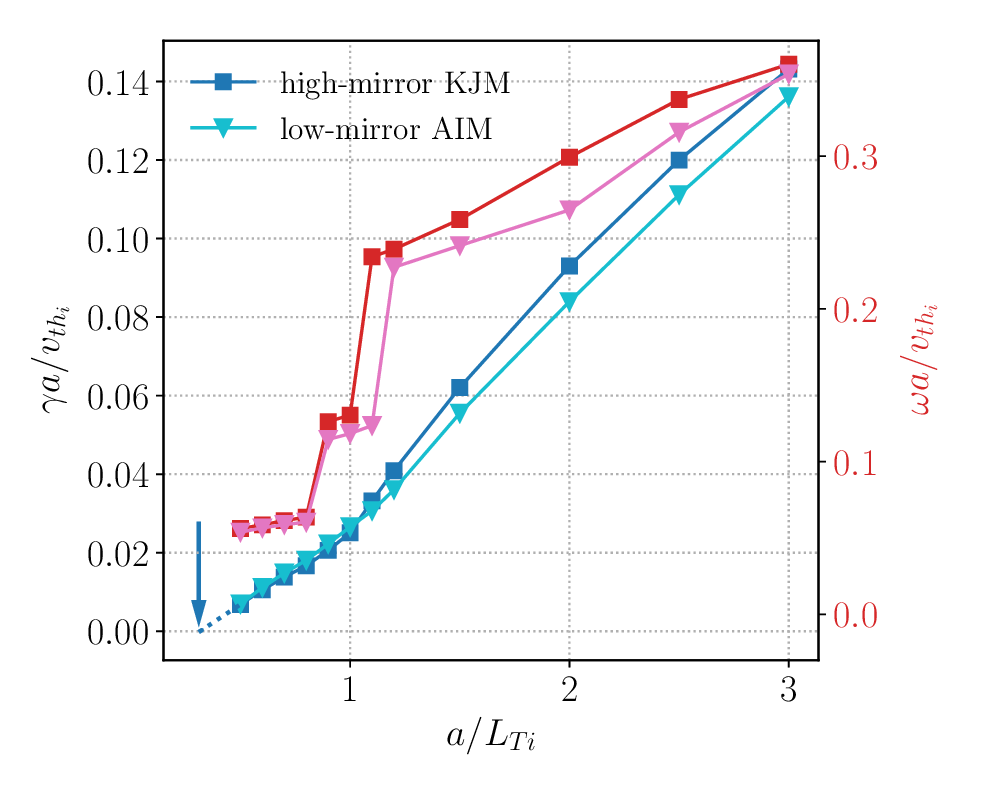}
    \caption{Normalised linear growth rates $\gamma a/v_{thi}$ -- in blue -- and frequencies $\omega a/v_{thi}$ -- in red -- for the fastest growing mode as a function of the ion temperature gradient showing the transition to Floquet background modes for the high-mirror and low-mirror configurations.}
    \label{fig:foot_new}
\end{figure}

The result is presented in Fig.~\ref{fig:spectra_new}, where plots of the normalised growth rate $\gamma a/v_{thi}$ and normalised frequency $\omega a/v_{thi}$ are shown. For each case we record the growth rate and the frequency of the fastest growing mode and plot these as functions of the normalised ion temperature gradient in Fig.~\ref{fig:foot_new}. We observe that for large ion temperature gradients, the fastest growing mode is found at the ion scale, $k_y\rho_i\approx 1$, consistently with a curvature-driven ITG. When the gradient is decreased, the curvature-driven ITG mode gets progressively stabilised, while other ``background'' modes emerge. The latter are characterised by lower values of $k_y\rho_i$ and are important only for sufficiently small values of the temperature gradient. We note that for a given $a/L_{Ti}$, different wavelengths can belong to different ITG branches. As a result, in the plot of the fastest-growing instability, there may be sudden jumps in the real part of the frequency. The background modes have been previously observed in low-shear tokamak and W7-X simulations featuring adiabatic electrons. Their nature can neither be attributed to the slab nor to the toroidal branch of the ITG mode. They are instead of the Floquet-type \citep{zocco2018threshold} and survive also when electrons are treated kinetically. A prominent feature of Floquet modes is their extended nature along magnetic field lines \citep{PhysRevE.106.L013202}. 

Blank spots in the spectra in Fig.~\ref{fig:spectra_new} represent values of $k_y\rho_i$ for which modes belonging to different branches are characterised by similar growth rates. In these cases, it is difficult to obtain a numerically converged spectrum. We thus plot the value of the growth rate and frequency by averaging over the last 10\% time steps of the simulation and by considering only points that are converged with a relative error of the order of a few percent.  

\begin{figure}
    \centering
    \includegraphics[scale=0.44]{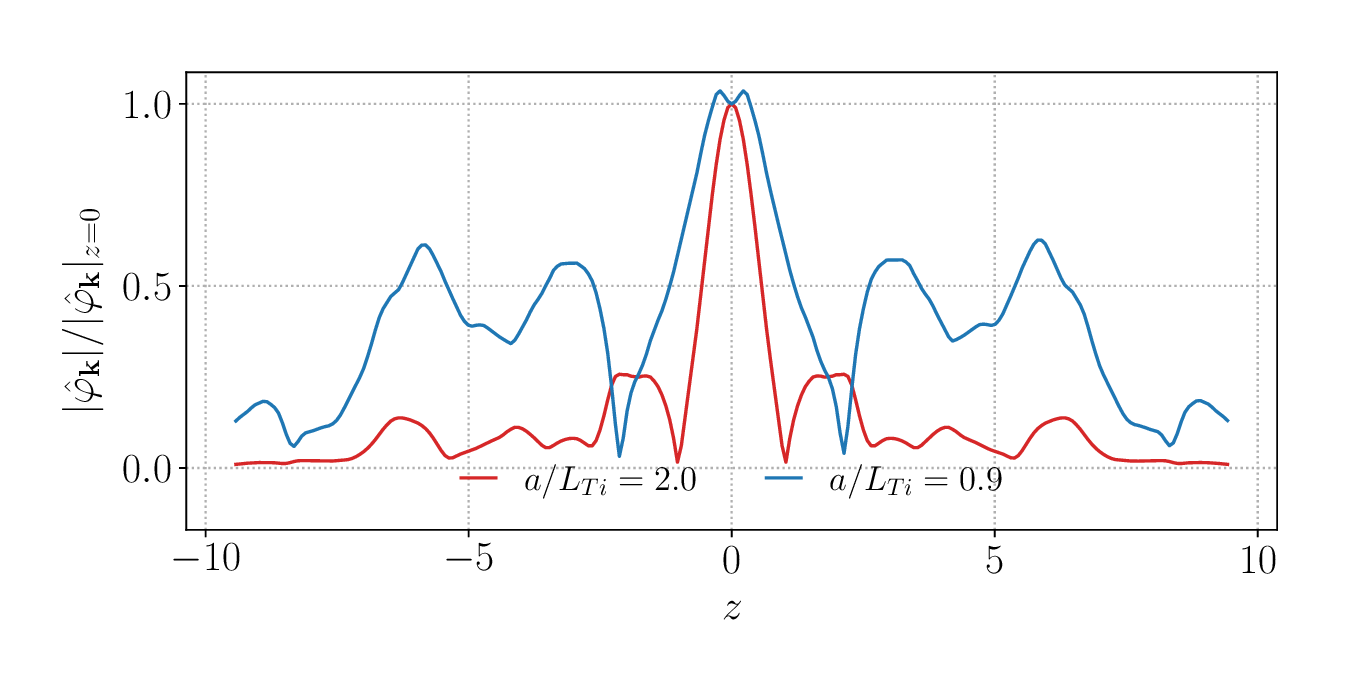}
    \caption{Fastest growing mode electrostatic potential structure along the field line in the fluid-like ($a/L_{Ti}=2.0$) case -- in red -- and in the Floquet-type ($a/L_{Ti}=0.9$) -- in blue.}
    \label{fig:eigenfunctions}
\end{figure}

In Fig.~\ref{fig:foot_new} we observe that the transition to Floquet-type modes is characterised by a change in the slope of the growth rate with the temperature gradient, which becomes less steep at small values of $a/L_{Ti}$, and by jumps in the frequency. The transition occurs between $a/L_{Ti}=1.0$ and $a/L_{Ti}=1.1$. In order to locate the linear stability threshold, we perform a linear interpolation of the last points corresponding to Floquet modes, and we find that the linear instability threshold is $a/L_{Ti}\lesssim 0.5$, which is consistent with what was found in \citep{zocco2018threshold, PhysRevE.106.L013202}. The linear results with kinetic electrons thus confirm the original findings obtained with adiabatic electrons. 
The same computation has been repeated considering the same input parameters but running the simulations in a W7-X low-mirror configuration (AIM). The result is also reported in Fig.~\ref{fig:foot_new}. We do not observe any particular dependence on the geometry, especially for the Floquet modes. The transition is visible in both geometries but is somewhat less pronounced in the low-mirror configuration. The values of the growth rates and frequencies tend to the same value as the gradient is progressively decreased. This suggests that such modes are relatively persistent and insensitive to the mirror ratio. 

\begin{figure}
    \centering
    \includegraphics[scale=0.45]{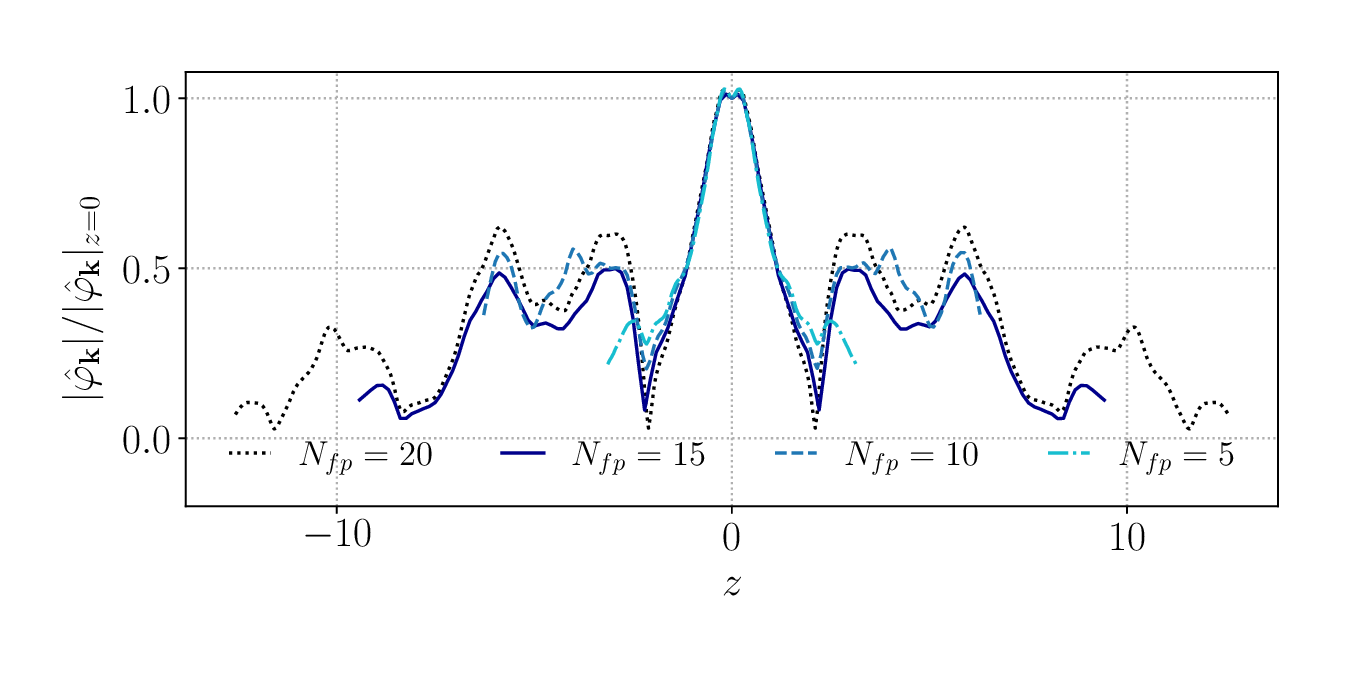}
    \caption{Comparison between electrostatic potential structures along the field line obtained with simulated flux-tubes of different lengths for the case $a/L_{Ti}=0.9$.}
    \label{fig:test_eigenfunctions}
\end{figure}

The transition between the two different instability branches is also visible from the change in the parallel structure of the electrostatic potential \citep{zocco2018threshold, PhysRevE.106.L013202}. To illustrate this fact, we choose two temperature gradients straddling the transition, $a/L_{Ti}=0.9$ and $a/L_{Ti}=2.0$. The first case corresponds to the Floquet-type ITG branch, while the second lies in the toroidal strongly-driven branch. The ITG destabilisation for the latter can be described with a simple fluid picture. For this reason, we will refer to it as fluid-like. We plot the modulus of the normalised electrostatic potential for the most unstable wave number divided by its value at $z=0$, $|\hat{\varphi}_{\vb{k}}|/|\hat{\varphi}_{\vb{k}}|_{z=0}$ in Fig.~\ref{fig:eigenfunctions}. We observe that, also in this case, the qualitative picture found with adiabatic electrons is unchanged: the Floquet-like eigenfunction ($a/L_{Ti}=0.9$) decays more slowly along the magnetic field line than the fluid-like one ($a/L_{Ti}=2.0$). The mode close to marginal stability thus requires a longer flux-tube (and more points along the field line) for its parallel structure to be sufficiently resolved. To choose the suitable flux-tube length we run scans of the \texttt{nfield\_periods} input parameter, $N_{fp}$, to set the number of magnetic periods covered by the flux-tube. One toroidal turn around W7-X corresponds to $N_{fp}=5$, since W7-X is a five-fold symmetric device. We compare the eigenfunction for the Floquet-type ITG mode in flux-tubes extending one, two, three and four times around the torus. The result in Fig.~\ref{fig:test_eigenfunctions} shows how at least two toroidal turns are necessary to sufficiently resolve the parallel structure of the electrostatic potential. With at least three toroidal turns a substantial decay of the eigenfunction towards the end of the domain is observed, while just one toroidal turn is not enough to resolve the slowly decaying structure of the mode. The choice of the flux-tube is thus of crucial relevance for obtaining sufficient energy injection, in particular in nonlinear simulations. For this reason, a follow-up discussion about suitable choices of input parameters in the parallel direction is given in the next section.

\section{Nonlinear results}
\label{sec:nonlinear}

\begin{table}
\begin{center}
\def~{\hphantom{0}}
\begin{tabular}{p{0.1\textwidth}*{6}{>{\centering\arraybackslash}p{0.1\textwidth}}}
	$\mathbf{a/L_{Ti}}$ & $\mathbf{N_z}$ & $\mathbf{N_{fp}}$ & $\mathbf{N_{v_{\parallel}}}$ & $\mathbf{N_{\mu}}$ & $\mathbf{N_{k_y}}$ & $\mathbf{N_{k_x}}$\\
    \hline
	2.0 & 96 & 5.6 & 24 & 12 & 151 & 91\\
	0.9 & 128 & 10 & 64 & 24 & 178 & 88\\
\end{tabular}
\caption{Set of input parameters for the nonlinear simulations in the fluid-like limit ($a/L_{Ti}=2.0$) and closer to marginality ($a/L_{Ti}=0.9$).}
\label{tab:resolutions}
\end{center}
\end{table}

We now move on to characterise differences between Floquet-type and fluid-like turbulence through nonlinear simulations using the same two values of the ion temperature gradient as in the previous section, $a/L_{Ti}=0.9$ and $a/L_{Ti}=2.0$. The input parameters are listed in table \ref{tab:resolutions}. Those for the simulation in the fluid-like regime ($a/L_{Ti}=2.0$) are chosen based on previous results \citep{garcia2021turbulent}; while, closer to marginality ($a/L_{Ti}=0.9$), the parameters are adjusted in accordance with the linear results mentioned above. Given the set of parameters displayed in the table, we are able to simulate the following wave numbers: in the fluid-like regime $k_{x,max}\rho_i\equiv\hat{k}_{x,max}\approx 1.19$ and $\hat{k}_{y,max}= 2.5$, with a respective step of $\Delta\hat{k}_x\approx 0.04$ and $\Delta\hat{k}_y= 0.05$, whereas closer to marginality: $\hat{k}_{x,max}\approx 2.06$ and $\hat{k}_{y,max}= 2.95$, with steps of $\Delta\hat{k}_x\approx 0.07$ and $\Delta\hat{k}_y= 0.05$. 

\begin{figure}
    \centering
    \includegraphics[scale=0.57]{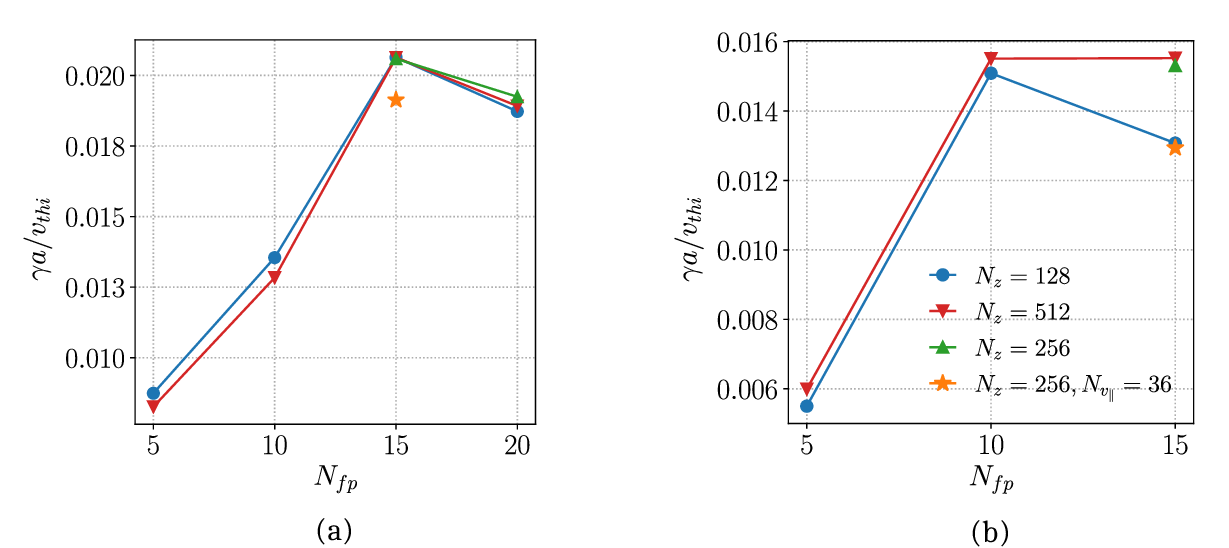}
    \caption{Flux-tube length convergence tests for growth rates at $a/L_{Ti} = 0.9$ for the wave numbers (a) $k_y\rho_i=0.71$ and (b) $k_y\rho_i=1.22$.}
    \label{fig:all_ky_convergence}
\end{figure}

We choose $N_{fp}=10$, i.e. two toroidal turns around the device, for the simulation close to marginality. This is different from what did linearly and the reason is computational. A longer flux-tube needs a larger number, $N_z$, of points in the $z$ direction, which makes the simulation computationally more expensive. We run convergence tests to ensure that two toroidal turns are enough to resolve the parallel structure of the Floquet modes and enough to obtain the energy injection needed by the instability. We first consider the most unstable wave number at $a/L_{Ti}=0.9$ and plot the value of the normalised growth rate as a function of the number of simulated field periods in Fig.~\ref{fig:all_ky_convergence}a. We see that with at least three toroidal turns we obtain a good convergence, with only a variation of the growth rate of 5\% when an additional toroidal turn is simulated. We also notice that $N_z=128$ is enough to resolve the parallel structure of the electrostatic potential. For $N_{fp}<15$, the value of the growth rate decreases by roughly 30\% for each toroidal turn that is removed from the simulation. When studying higher values of $k_y\rho_i$, as in Fig.~\ref{fig:all_ky_convergence}b, we notice that a larger value of $N_z$ is required to obtain convergence for $N_{fp}=15$. For computational reasons, however, we settle on two toroidal turns, i.e. $N_{fp}=10$, with $N_z=128$, so as to obtain a less computationally expensive simulation but yet enough resolution for a sensible result. In fact, even if the wave numbers around $k_y\rho_i\approx 1$ are not the most unstable ones close to marginality, they still play a role and need to be properly resolved too. In the two plots, the importance of having enough parallel velocity space resolution is also evident.

\begin{figure}
    \centering
    \includegraphics[scale=0.53]{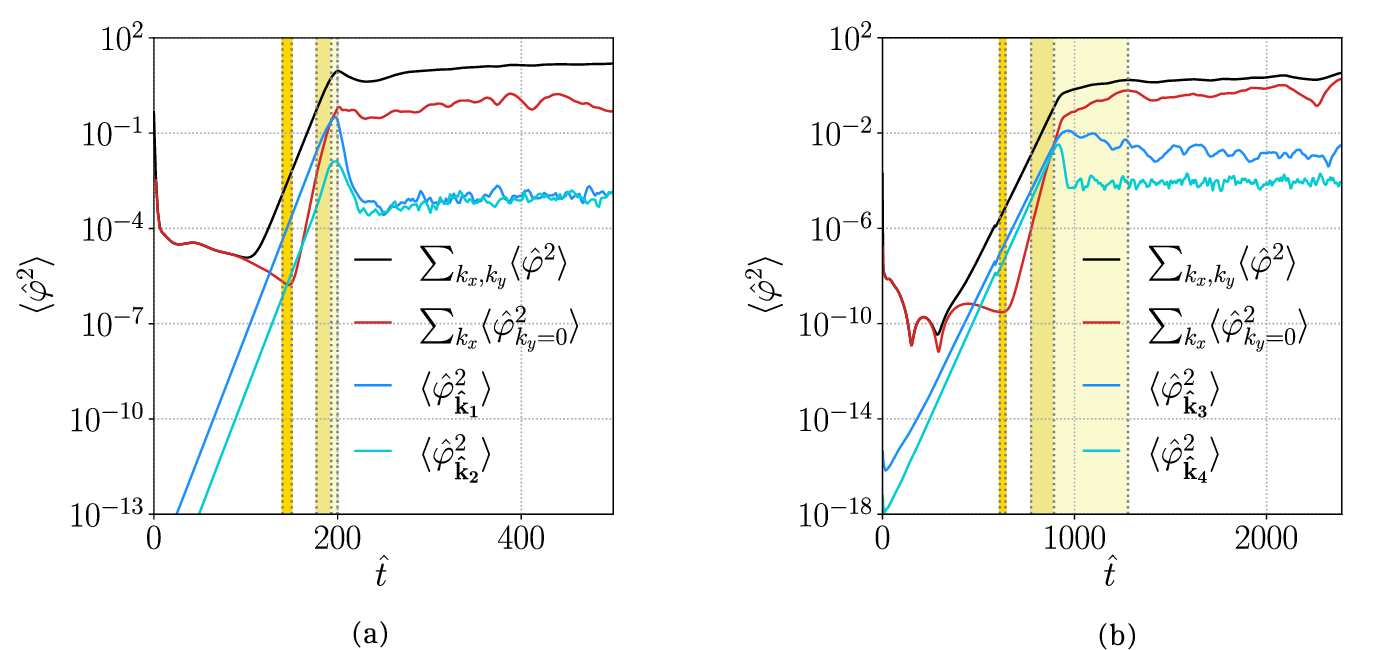}
    \caption{Time evolution of the average squared modulus of the electrostatic potential fluctuations $\langle \hat{\varphi}^2\rangle$ -- in black -- compared to its zonal component $\sum_{k_x}\langle\hat{\varphi}^2_{k_y=0}\rangle$ -- in red -- and the primaries -- in blue -- for (a) $a/L_{Ti}=2.0$ and (b) $a/L_{Ti}=0.9$. Time windows encompassing the three key time instants discussed in the text are displayed as yellow bands. Adapted from \citep{PhysRevE.106.L013202}.}
    \label{fig:all_time_traces}
\end{figure}

In \citep{PhysRevE.106.L013202}, the evolution of the electrostatic potential fluctuations was already studied in order to characterise the turbulence. We pay particular attention to the average squared modulus $\langle \hat{\varphi}^2_{\vb{k}} \rangle$ as a function of the normalised time $\hat{t}=tv_{thi}/a$. The time trace for $a/L_{Ti}=2.0$ is displayed in Fig.~\ref{fig:all_time_traces}a, and the one for $a/L_{Ti}=0.9$ in Fig.~\ref{fig:all_time_traces}b. We specifically compare the time evolution of the zonal structures ($\hat{k}_y=0$) with the one of the primary linear instabilities. The relative importance of the two gives rise to the definition of three time instants. At the first instant, zonal structures reach a minimum, after a transient phase. At the second, the amplitude of zonal flows equals the one of the main linear instabilities, and at the third one, the electrostatic potential resulting from the summation of all the Fourier modes, $\langle \hat{\varphi}^2 \rangle=\sum_{k_x,k_y}\hat{\varphi}_{\vb{k}}$ reaches a maximum and saturates. 

\begin{figure}
    \centering
    \includegraphics[scale=0.57]{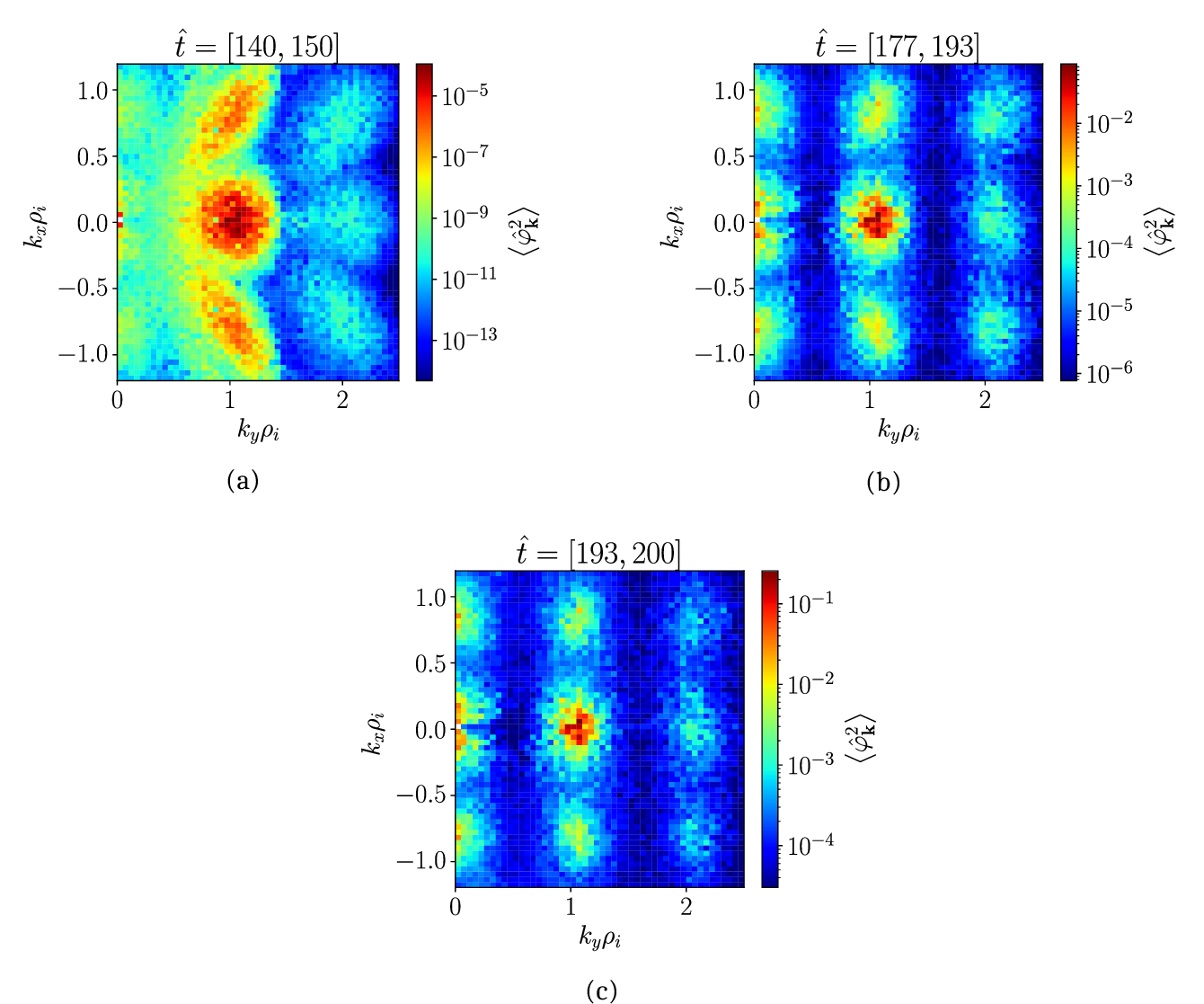}
    \caption{Fourier spectra of time averaged $\langle\hat{\varphi}^2_{\vb{k}}\rangle$ as a function of the normalised binormal $k_y\rho_i$ and radial $k_x\rho_i$ wave numbers for $a/L_{Ti}=2.0$ at the three key time instants: (a) first time instant -- adapted from \citep{PhysRevE.106.L013202}, (b) second time instant and (c) third time instant.}
    \label{fig:all_tprim2_spectra}
\end{figure}

We study the Fourier spectra of the fluctuations for each time instant, performing an average over a time window about each instant. We start with the fluid-like scenario ($a/L_{Ti}=2.0$). In the first time instant (Fig.~\ref{fig:all_tprim2_spectra}a) the primary linear instability can be studied. It consists of an isotropic structure at $\vb{k}_1\equiv(\hat{k}_x,\hat{k}_y)=(0,1.05)$, typically observed in axisymmetric simulations, and two additional lobes at $\vb{k}_2\equiv(\hat{k}_x,\hat{k}_y)=(\pm 0.84,1.05)$. At the second instant, the evolution of the primary modes starts to depart from the linear phase and the zonal flows play a more prominent role. This can be considered as a pre-saturation phase. In this phase, the finite-$\hat{k}_x$ lobes get isotropised (Fig.~\ref{fig:all_tprim2_spectra}b) and their relative importance decreases. A checkerboard-like structure forms and persists also in the saturating phase (Fig.~\ref{fig:all_tprim2_spectra}c). From the time evolution in Fig.~\ref{fig:all_time_traces}a, we observe that after the saturating phase both the primary isotropic lobe and the finite-$\hat{k}_x$ lobes are suppressed by up to four orders of magnitude, while zonal flows still remain active.

\begin{figure}
    \centering
    \includegraphics[scale=0.57]{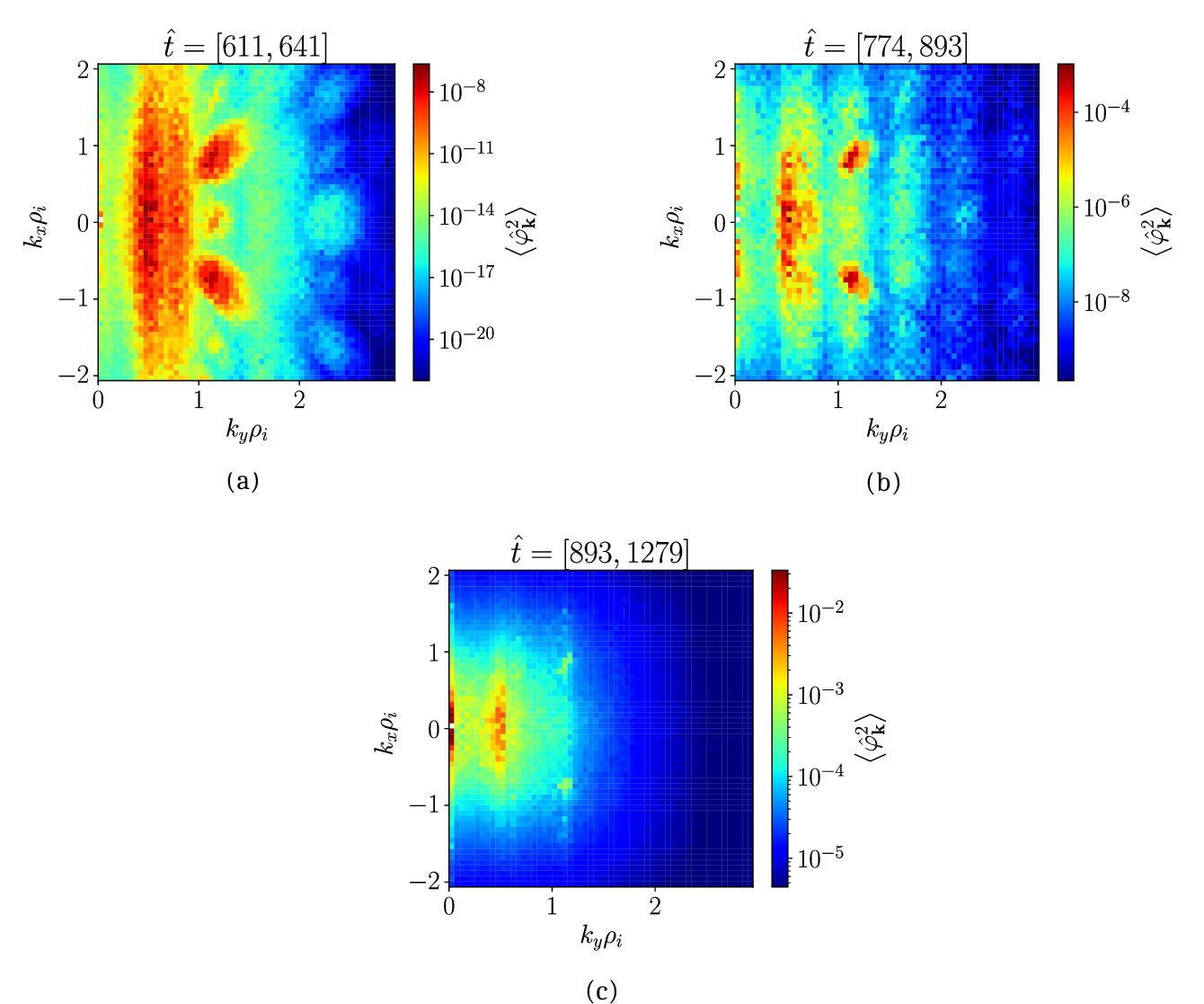}
    \caption{Fourier spectra of time averaged $\langle\hat{\varphi}^2_{\vb{k}}\rangle$ as a function of the normalised binormal $k_y\rho_i$ and radial $k_x\rho_i$ wave numbers for $a/L_{Ti}=0.9$ at the three key time instants: (a) first time instant -- adapted from \citep{PhysRevE.106.L013202}, (b) second time instant and (c) third time instant.}
    \label{fig:all_tprim09_spectra}
\end{figure}

Floquet-type turbulence ($a/L_{Ti}=0.9$) behaves differently but the first instant features similarities with the fluid-like case. In particular, both the isotropic structure and the finite-$\hat{k}_x$ lobes are present but are shifted to higher $\hat{k}_y$ (Fig.~\ref{fig:all_tprim09_spectra}a). The isotropic structure is also reduced in relative intensity compared with the finite-$\hat{k}_x$ lobes. However, they do not appear as the primary in this case. The primary mode develops around $\hat{k}_y\approx 0.5$ and shows a broader structure in $\hat{k}_x$. This confirms what was found linearly (Fig.~\ref{fig:spectra_new}) and suggests that turbulence might have a narrower radial structure closer to marginality. This new primary mode develops in a region that is stable in simulations at higher temperature gradients.

The time evolution of the primaries is different from the fluid-like case. Both the primary characterised by the band in $\hat{k}_x$ and the two finite-$\hat{k}_x$ lobes persist in the pre-saturation phase, while the isotropic structure decreases in relative intensity (Fig.~\ref{fig:all_tprim09_spectra}b). Their time evolution is shown in Fig.~\ref{fig:all_time_traces}b. In particular, $\vb{k}_3\equiv(\hat{k}_x,\hat{k}_y)=(0.14,0.5)$ and $\vb{k}_4\equiv(\hat{k}_x,\hat{k}_y)=(0.71,1.05)$ are respectively chosen as representative. We note that the $\hat{k}_x$ band shows a longer decaying time and saturates at higher values compared to the finite-$\hat{k}_x$ lobe. This feature is also visible from the Fourier spectra of fluctuations at the third time instant (Fig.~\ref{fig:all_tprim09_spectra}c) and is attributed to a reduced efficiency of the zonal flows in suppressing the broad $\hat{k}_x$ band.

\begin{figure}
    \centering
    \includegraphics[scale=0.45]{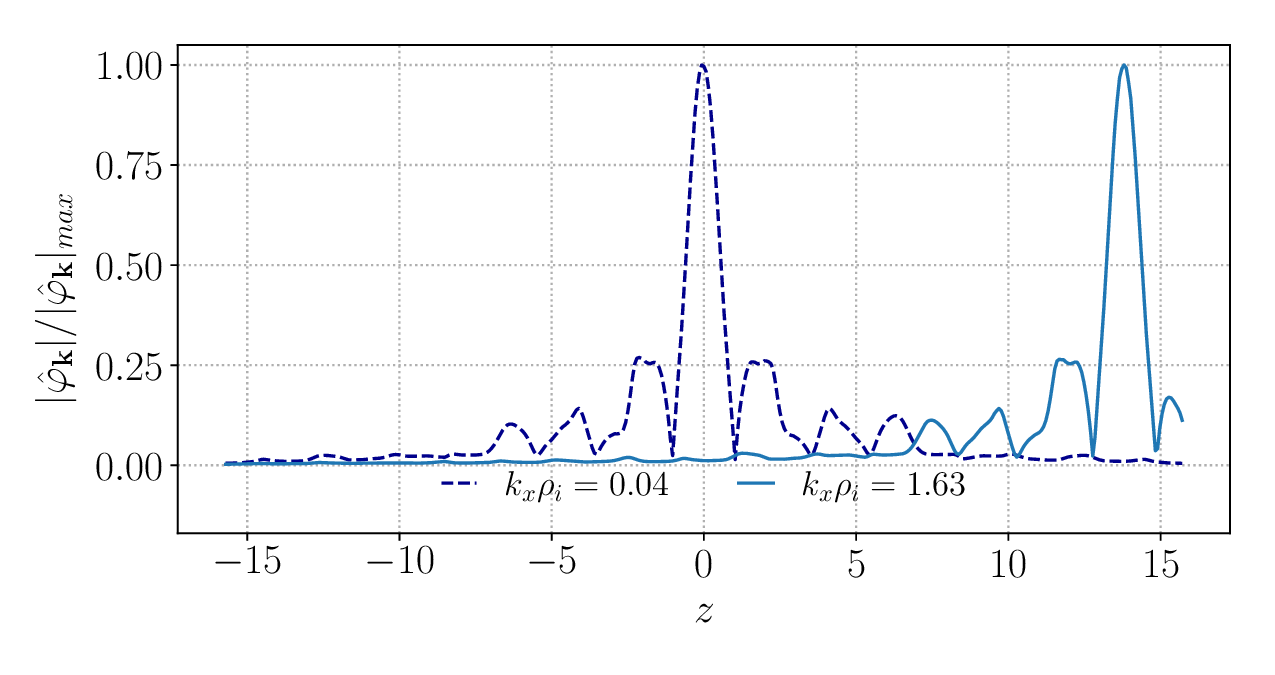}
    \caption{Electrostatic potential structure along the magnetic field line for $a/L_{Ti}=2.0$ at the binormal wave number $k_y\rho_i=1.05$ and two different radial wave numbers: $k_x\rho_i=0.04$ and $k_x\rho_i=1.63$.}
    \label{fig:eigenf_nfp25}
\end{figure}

\begin{figure}
    \centering
    \includegraphics[scale=0.41]{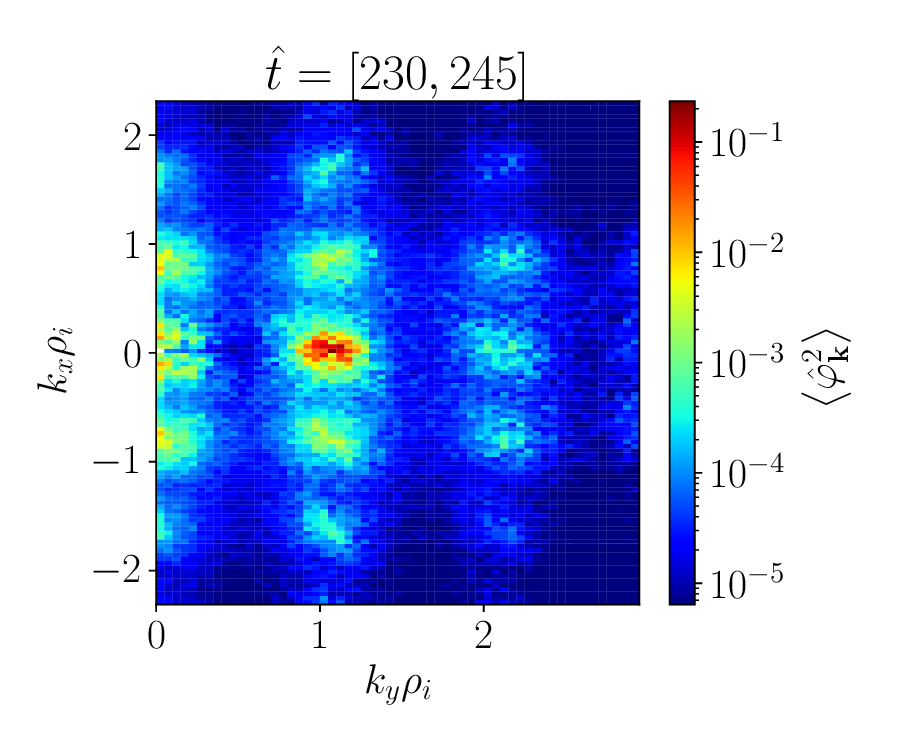}
    \caption{Fourier spectrum of $\langle\hat{\varphi}^2_{\vb{k}}\rangle$ as a function of the binormal $k_y\rho_i$ and radial $k_x\rho_i$ wave numbers during the pre-saturation phase for $a/L_{Ti}=2.0$. Simulation with a larger range of $k_x\rho_i$.}
    \label{fig:spectrum_largekx}
\end{figure}

The finite-$\hat{k}_x$ lobes, present in both the fluid-like and the Floquet-type cases, are understood as modes that peak further along the magnetic field line -- see Fig.~\ref{fig:eigenf_nfp25}. By running a simulation for $a/L_{Ti}=2.0$ with a larger $k_x\rho_i$ domain, we observe that the checkerboard-like structure extends even more in $k_x\rho_i$ (Fig.~\ref{fig:spectrum_largekx}). However, the intensity of the lobes decreases with increasing $k_x\rho_i$ and the saturated level of fluctuations remains unchanged. We can conclude that there is no need to run larger simulations to include the whole checkerboard-like structure.

\begin{figure}
    \centering
    \includegraphics[scale=0.5]{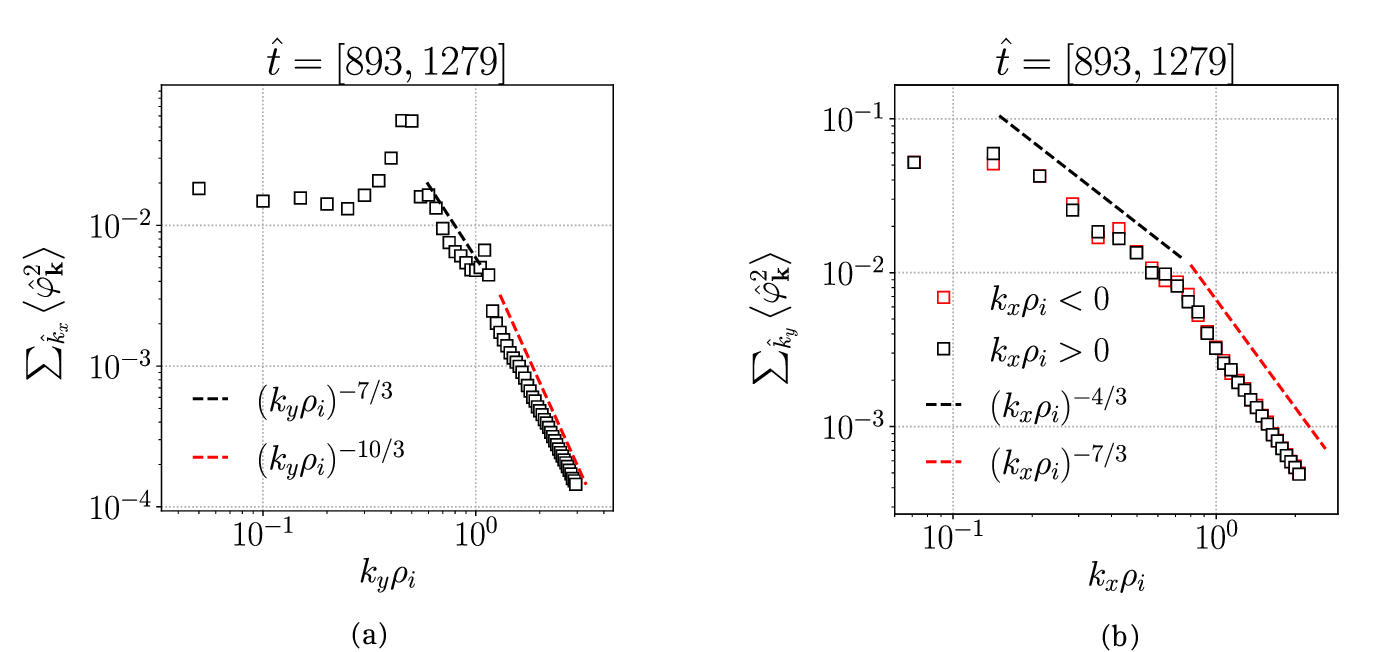}
    \caption{Electrostatic fluctuations spectra for $a/L_{Ti}=0.9$ in (a) $k_y\rho_i$ and (b) $k_x\rho_i$ at the third time instant and compared with the predicted scalings. Adapted from \citep{PhysRevE.106.L013202}.}
    \label{fig:all_scalings}
\end{figure}

At the third time instant, marginal turbulence in W7-X sets up a steep sub-$\rho_i$ inertial range (Fig.~\ref{fig:all_scalings}a). A transition at the ion Larmor radius is observed \citep{PhysRevE.106.L013202}, as for $k_y\rho_i\lesssim 1$ (long wavelenghts) the energy cascade is consistent with a $\sum_{k_x}\langle\hat{\varphi}^2_{\vb{k}}\rangle\sim (k_y\rho_i)^{-7/3}$ scaling \citep{critical}, while for $k_y\rho_i\gtrsim 1$ the sub-$\rho_i$ $\sum_{k_x}\langle\hat{\varphi}^2_{\vb{k}}\rangle\sim (k_y\rho_i)^{-10/3}$ scaling \citep{Schekochihin_2008} is observed. Both these scalings were observed before \citep{gabesat} but the former was suggested to be characteristic of W7-X, and the latter of the quasihelically symmetric stellarator HSX. While the long-wavelength $(k_y\rho_i)^{-7/3}$ scaling is only observed over a limited inertial range, the sub-$\rho_i$ $(k_y\rho_i)^{-10/3}$ one persists for over a decade in $k_y\rho_i$, allowing a quantitative interpolation $\sum_{k_x}\langle\hat{\varphi}^2_{\vb{k}}\rangle\sim (k_y\rho_i)^{-\alpha/3},$ yielding $\alpha=10.1\pm0.3$. A similar transition, at ion Larmor radius scales, is observed in the radial spectrum of fluctuations (Fig.~\ref{fig:all_scalings}b). However, there is no explanation for such scalings at the present time.

\begin{figure}
    \centering
    \includegraphics[scale=0.35]{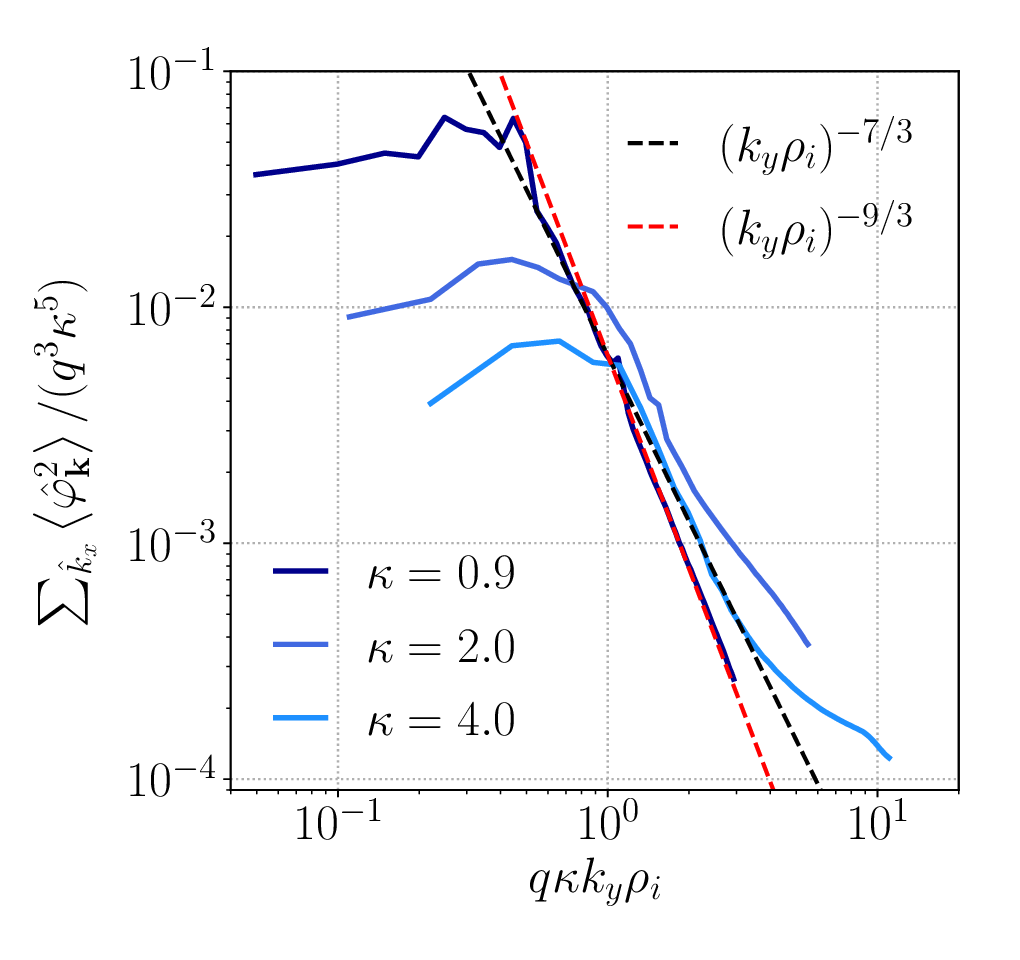}
    \caption{Electrostatic fluctuations spectra for three different ion temperature gradients as a function of $k_y\rho_i$ and normalised as in \citep{critical}. The fully developed turbulent regime is considered and compared with predicted scalings. $q$ is the safety factor and $\kappa=a/L_{Ti}$.}
    \label{fig:scalings_comparison}
\end{figure}

To understand the energy cascade at the third instant, it is useful to consider the primary linear source of turbulence that needs to be suppressed by zonal flows. We also study the energy cascade at later instants, when turbulence is fully developed, i.e., at $\hat{t}>400$ for $a/L_{Ti}=2.0$ and $\hat{t}>1279$ for $a/L_{Ti}=0.9$. In Fig.~\ref{fig:scalings_comparison} we have plotted the energy spectrum for three different values of the ion temperature gradient, normalised as in \citep{critical}. The normalisation amounts to a stretching of the inertial range where the $(k_y\rho_i)^{-7/3}$ trend is expected to be observed. Since the normalisation factors include an $a/L_{Ti}$ factor, such stretching is larger for larger ion temperature gradients. The range over which the $(k_y\rho_i)^{-7/3}$ scaling is observed is the largest for $a/L_{Ti}=4.0$ and it decreases with decreasing $a/L_{Ti}$. Hence the reduced inertial range in which the $(k_y\rho_i)^{-7/3}$ scaling is observed for $a/L_{Ti}=0.9$. The transition to a sub-$\rho_i$ steeper scaling is still observed but it is characterised by a less pronounced slope, which is, however, incompatible with any scaling shallower than $(k_y\rho_i)^{-9/3}$. The observation of the sub-$\rho_i$ transition can thus be put in a broader context of a general steepening of turbulence spectra with decreasing ion temperature gradient \citep{PhysRevE.106.L013202}. 

\begin{figure}
    \centering
    \includegraphics[scale=0.5]{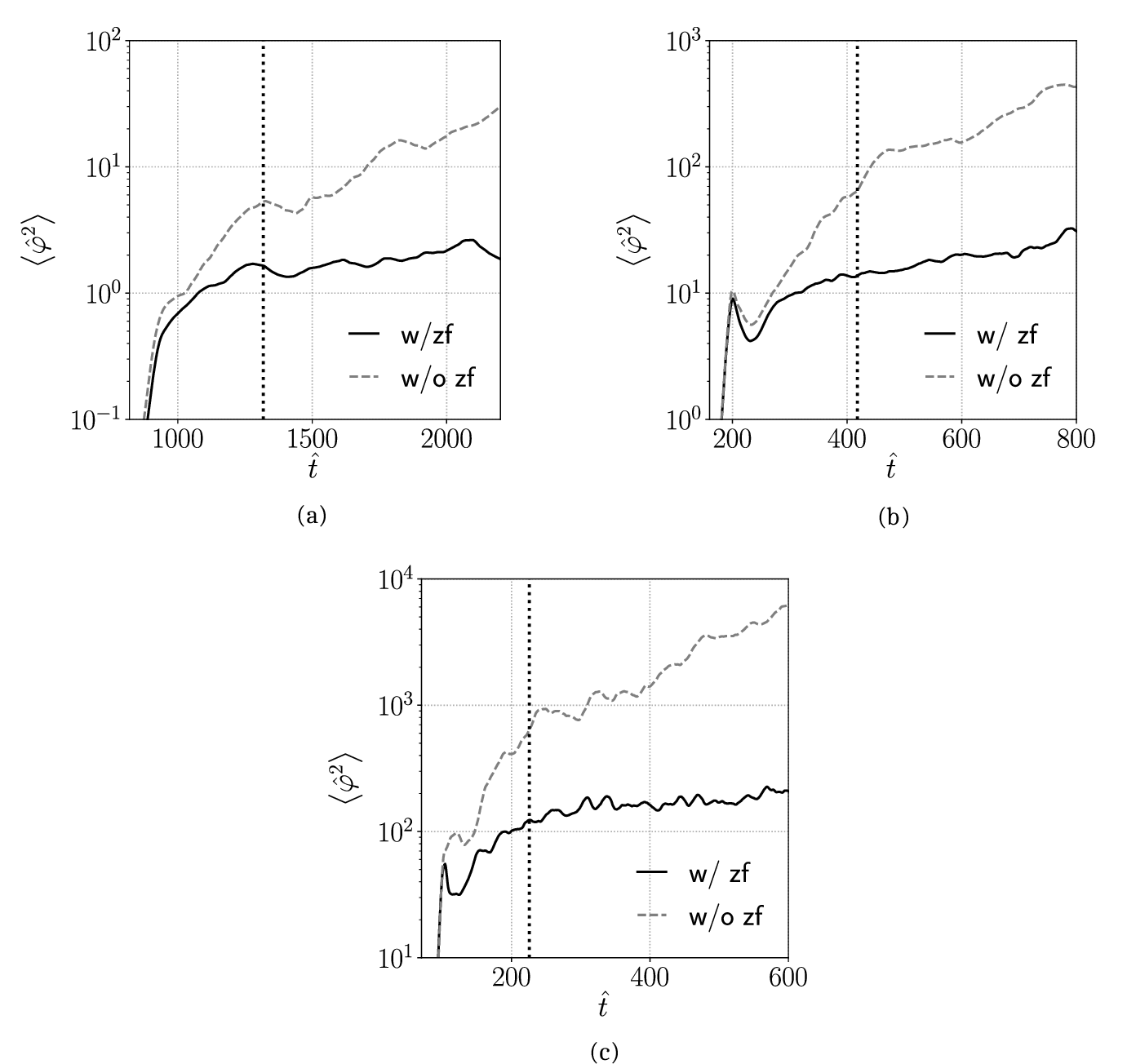}
    \caption{Time evolution of the average squared modulus of the electrostatic potential fluctuations $\langle \hat{\varphi}^2\rangle$ for simulations with and without zonal flows in the $\vb{E}\times\vb{B}$ nonlinearity: (a) $a/L_{Ti}=0.9$, (b) $a/L_{Ti}=2.0$ and (c) $a/L_{Ti}=4.0$.}
    \label{fig:all_no_zf}
\end{figure}

We have previously observed and suggested that turbulence close to marginality in W7-X is more radially localised and thus less prone to nonlinear zonal flows suppression \citep{PhysRevE.106.L013202}. The mechanism that allows such suppression is the shearing of turbulent structures by zonal flows, which causes an up-shift of the nonlinear critical gradient with respect to the linear stability threshold. This up-shift is known as Dimits shift \citep{rogers, st-onge_2017} and even though in tokamaks for realistic conditions \citep{mikkelsendorland} it can be less abrupt than originally observed \citep{dimitz}, it is always present. In order to explore whether this is also true in W7-X, we run further simulations where we artificially suppress zonal flows in the $\vb{E}\times\vb{B}$ nonlinearity. Results of such simulations are reported in Figs.~\ref{fig:all_no_zf}a, \ref{fig:all_no_zf}b and \ref{fig:all_no_zf}c. The zonal flows suppression causes the electrostatic fluctuations to grow faster and to saturate at larger amplitudes than before. The result is compatible with what is observed in tokamaks \citep{Lin1835}, but the effect is not as strong. In all three cases the amplitude of electrostatic fluctuations is enhanced, but when comparing the value of fluctuations case by case, we see that the effect decreases with decreasing $a/L_{Ti}$. To systematically analyse the simulations, we consider an instant just after the pre-saturation phase (identified by the dotted lines in the plots) and we calculate the ratio $r(t)=\langle\hat{\varphi}^2_{\text{w/o}}\rangle/\langle\hat{\varphi}^2_{\text{w/}}\rangle$. For the set of gradients $a/L_{Ti}=\{0.9,2.0,4.0\}$ we find that $r=\{3.3,4.7,5.1\}$. The ratio thus decreases with decreasing temperature gradient suggesting that the zonal flows are less effective when the temperature gradient is small \citep{PhysRevE.106.L013202}.

\begin{figure}
    \centering
    \includegraphics[scale=0.6]{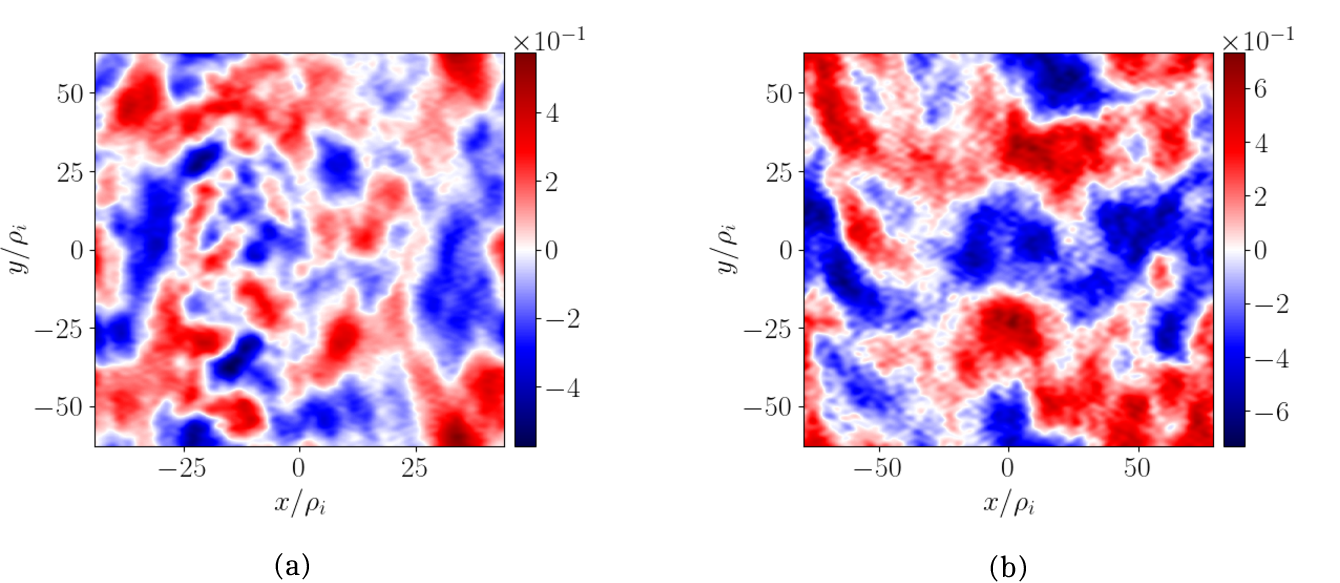}
    \caption{Real-space extent of electrostatic potential fluctuations in the saturated regime for (a) $a/L_{Ti}=0.9$ and (b) $a/L_{Ti}=2.0$.}
    \label{fig:all_real}
\end{figure}

To assess whether this reduced effectiveness is related to the spatial distribution of the turbulence, we compute the inverse Fourier transform of the non zonal components of $\langle\hat{\varphi}^2\rangle$. Figs.~\ref{fig:all_real}a and \ref{fig:all_real}b show the the real-space structure of fluctuations, in the saturated regime, for both fluid-like and Floquet-type turbulence. Note that the $x$ axes of the two figures have different scales due to the choice of different radial wave number resolutions. We observe that fluid-like turbulence is characterised by larger radial structures, with radial dimensions of more than $50 \rho_i$. Going towards marginality, we observe, as predicted, narrower radial structures. The fluctuations are generally more filamented and characterised by dimensions of $\approx 25 \rho_i$.   

\begin{figure}
    \centering
    \includegraphics[scale=0.37]{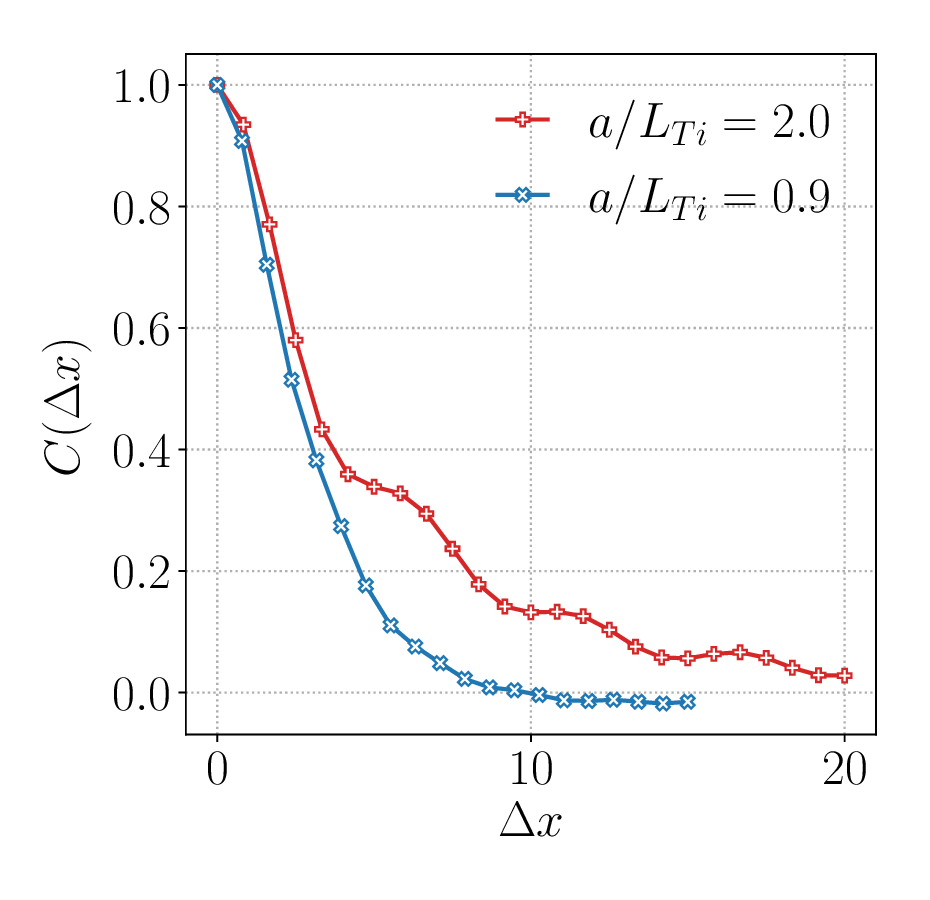}
    \caption{Turbulence radial correlation function in the fully developed turbulent regime for the fluid-like $(a/L_{Ti}=2.0)$ and Floquet-type turbulence $(a/L_{Ti}=0.9)$. Adapted from \citep{PhysRevE.106.L013202}.}
    \label{fig:corr_func}
\end{figure}

To more quantitatively characterise the radial extent of turbulent structures, we calculate the radial correlation function \citep{PhysRevE.106.L013202}
$$\mathcal C (\Delta x)=\left(\int \dd t\dd x\dd y\dd z|\varphi|^2\right)^{-1}\int \dd t\dd x\dd y\dd z \varphi(x)\varphi(x+\Delta x)$$
and the average radial correlation length 
$$\overline{\Delta x}=\int_0^{\infty} \mathcal C(\Delta x) \dd (\Delta x).$$
From Fig.~\ref{fig:corr_func} we see that the radial correlation function for Floquet-type turbulence decays faster with $\Delta x$ than the one for fluid-like turbulence. This is reflected in the values of the average radial correlation length, which for $a/L_{Ti}=\{2.0, 0.9\}$ are $\overline{\Delta x}=\{5.0, 2.9\}$.

\begin{figure}
    \centering
    \includegraphics[scale=0.4]{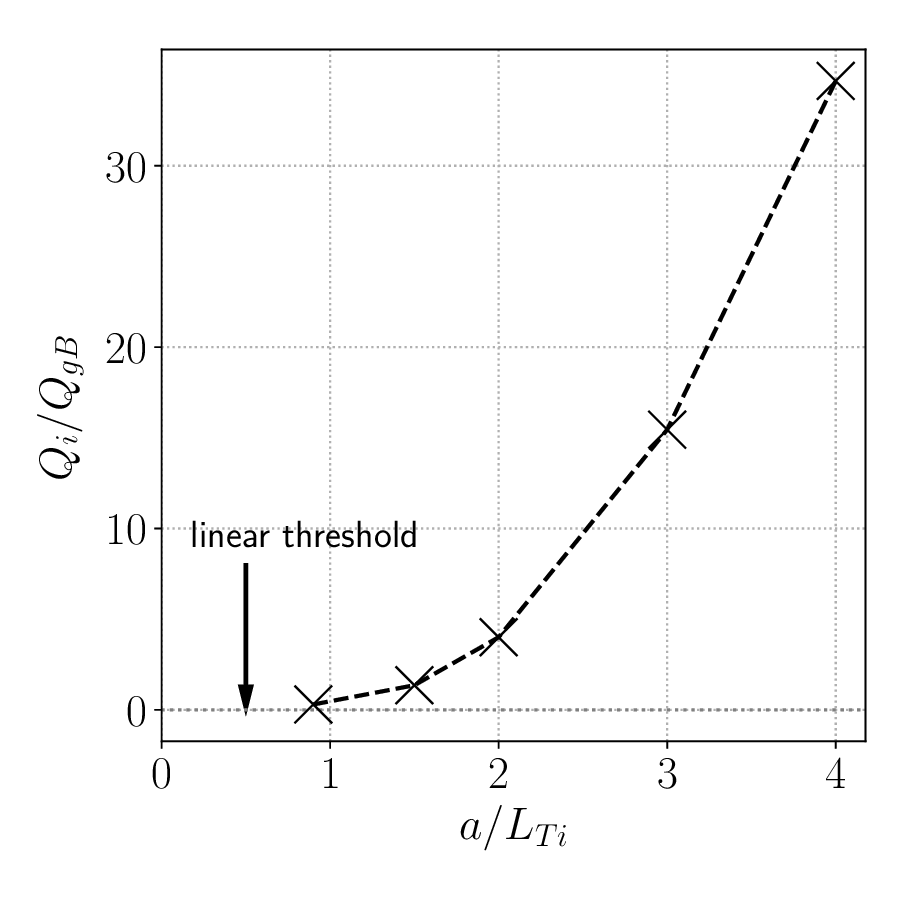}
    \caption{Normalised ion energy flux $Q_i/Q_{gB}$ as a function of the ion temperature gradient $a/L_{Ti}$ for different nonlinear simulations. The linear threshold is also indicated. Adapted from \citep{PhysRevE.106.L013202}.}
    \label{fig:Qi_vs_aLTi}
\end{figure}

\begin{figure}
    \centering
    \includegraphics[scale=0.45]{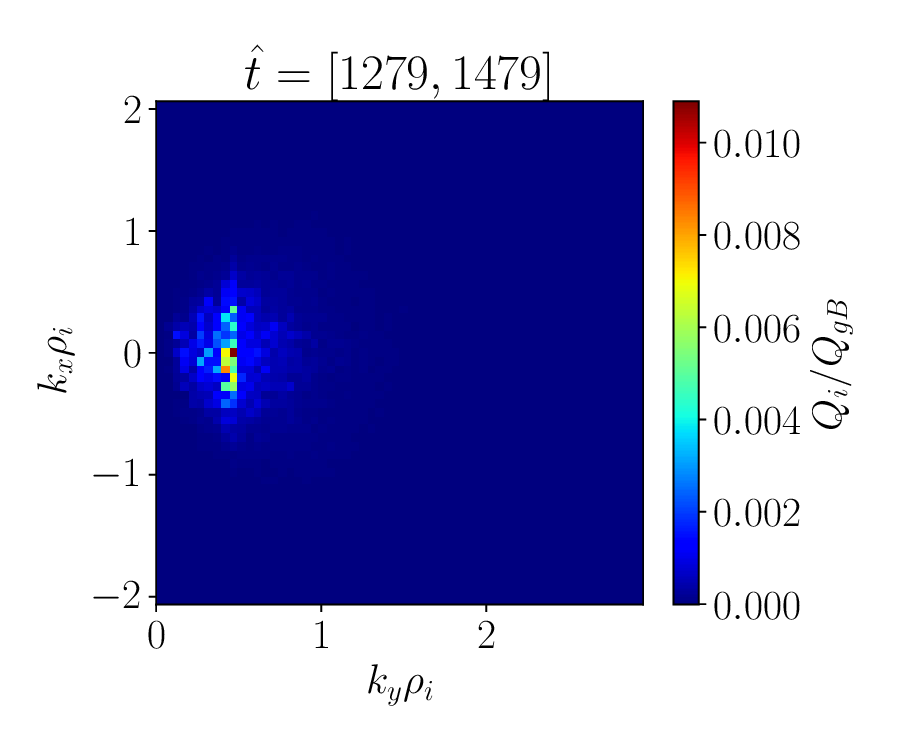}
    \caption{Contribution of various wave numbers to the normalised ion energy flux $Q_i/Q_{gB}$ in the fully developed turbulent regime for $a/L_{Ti}=0.9$. Adapted from \citep{PhysRevE.106.L013202}.}
    \label{fig:Qi_spectrum}
\end{figure}

These results suggest a reason for the lack of turbulence suppression when the temperature gradient is close to the linear stability threshold. Because of the small radial correlation length of Floquet-type turbulence, zonal flows are relatively ineffective, and a finite energy flux persists close to the stability threshold \citep{PhysRevE.106.L013202}. This can be observed in Fig.~\ref{fig:Qi_vs_aLTi}, showing the normalised ion energy flux versus the ion temperature gradient. Floquet-type turbulence produces a normalised energy flux of $Q_i/Q_{gB}\approx 0.3$. In order to obtain this plot, additional nonlinear simulations were performed at $a/L_{Ti}=\{1.5,3.0,4.0\}$. Note that there is a smooth transition from strongly-driven fluid-like generated transport to the Floquet-type one. The linear threshold is also indicated, and there is no sign of a Dimits shift or a nonlinear threshold for the onset of transport in W7-X. Similar behaviour has earlier been observed in turbulence simulations of quasihelical symmetric stellarators \citep{bader, terry_threshold} but is very different from that in tokamaks. Fig.~\ref{fig:Qi_spectrum} shows the contribution of different wave numbers to the turbulent energy flux for $a/L_{Ti}=0.9$. A comparison with Fig.~\ref{fig:all_tprim09_spectra}c confirms they belong to the broad $\hat{k}_x$-band characteristic of Floquet-like turbulence \citep{PhysRevE.106.L013202}.

\section{Shear effects on the ITG destabilisation}
\label{sec:shear}

\begin{figure}
    \centering
    \includegraphics[scale=0.45]{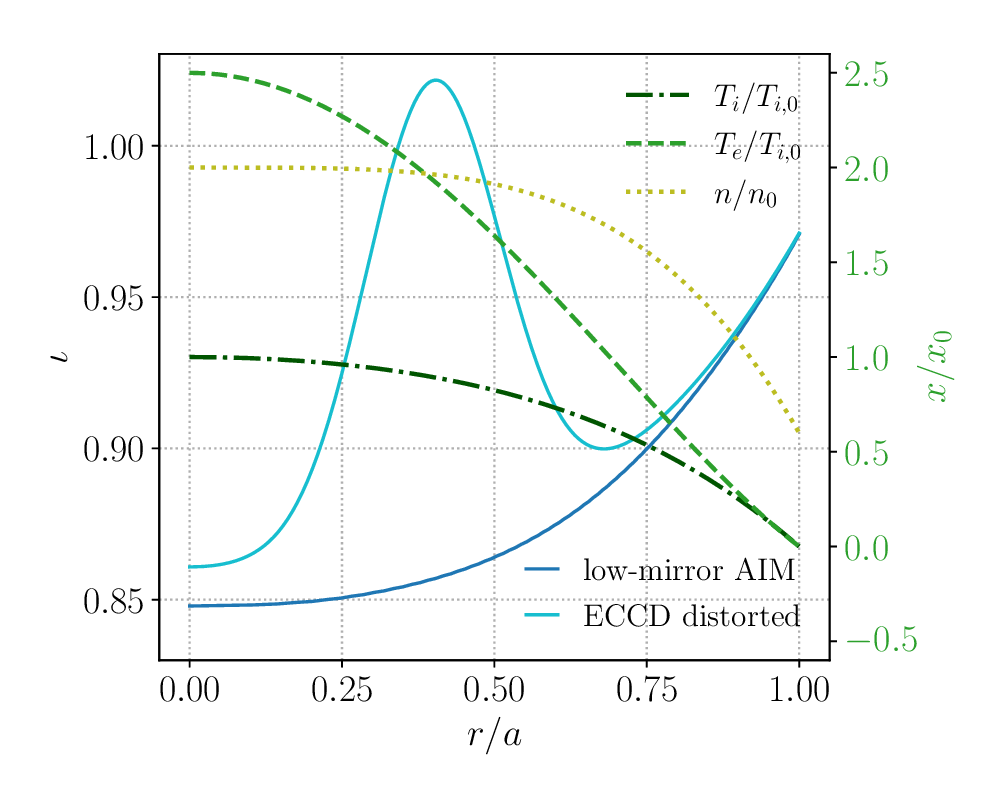}
    \caption{$\iota$ profiles for the low-mirror and ECCD non-vacuum distorted low-mirror equilibrium  -- in blue. In green, the normalised ion and electron temperature and density profiles used to construct the non-vacuum equilibrium.}
    \label{fig:iota_profiles}
\end{figure}

In this section we study the effects of global magnetic shear $\hat{s}=-(r/\iota) \dd \iota/\dd r$ on ITG destabilisation in W7-X, with an emphasis on Floquet modes and the instability threshold. We are particularly interested in exploring the effects of a strong and positive magnetic shear. W7-X usually possesses a low, negative shear (in the tokamak sense, $\hat s > 0)$ but can acquire larger shear of either sign if a significant toroidal current is present in the plasma. Such a current arises in practice through electron cyclotron current drive (ECCD), which tends to create a local maximum in the $\iota$-profile, with regions of significant shear on either side of the maximum \citep{zanini2020eccd, zanini2021confinement, aleynikova2021model, Zocco_2021}.  Accordingly, we proceed from an MHD equilibrium computed from the vacuum low-mirror (AIM) W7-X configuration, with prescribed profiles of density, temperature and rotational transform shown in Fig.~\ref{fig:iota_profiles}. 

\begin{figure}
    \centering
    \includegraphics[scale=0.67]{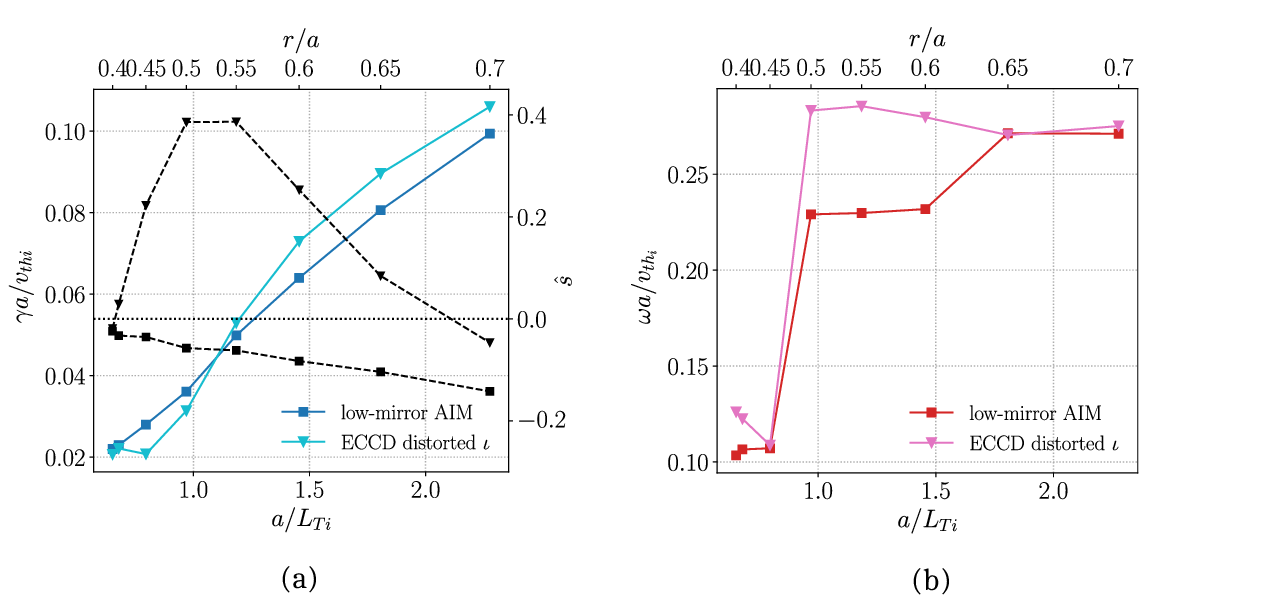}
    \caption{Normalised linear growth rates in (a) and normalised frequencies in (b), both as a function of the ion temperature gradient and radial position for the low-mirror configuration and ECCD distorted $\iota$ equilibrium. In (a) the radial profile of the global shear $\hat{s}$ is plotted in dashed black.}
    \label{fig:foot_bumpyiota}
\end{figure}

In the magnetic geometry of this equilibrium, we run linear, flux-tube, electrostatic simulations with kinetic electrons and similar resolutions to those above. Each simulation is carried out at a different radial position $r/a$, mostly in the positive-shear region. The radial scan implies a scan in the shear $\hat{s}$ and in the normalised ion temperature gradient $a/L_{Ti}$. We also change $\tau$ consistently with the prescribed plasma profiles, but we set $a/L_{Te}=a/L_n=0$. We compare the simulations performed in the ECCD distorted equilibrium with the ones obtained with the original low-mirror equilibrium without plasma current. We summarise in table \ref{tab:simulations} in Appendix \ref{sec:appendix} the radial positions where the simulations were carried out, the corresponding values of the local shear for the two equilibria ($\hat{s}_{\text{ECCD}}$ and $\hat{s}_{\text{AIM}}$), the normalised ion temperature gradient $a/L_{Ti}$, and the temperature ratio $\tau$, together with the resulting growth rates. 

\begin{figure}
    \centering
    \includegraphics[scale=.9]{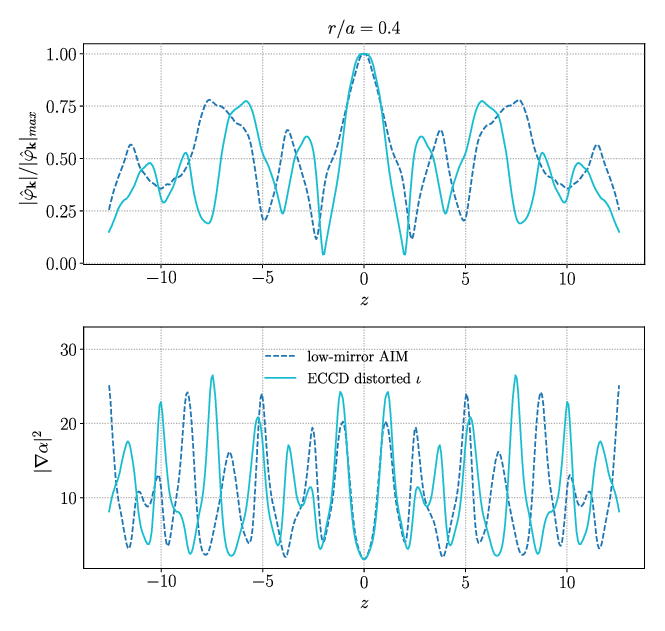}
    \caption{Comparison of the electrostatic potential and $|\nabla\alpha|^2$ structure along the magnetic field line for the low-mirror equilibrium -- dashed line -- and the ECCD distorted $\iota$ one -- solid line -- at $r/a=0.4$.}
    \label{fig:AIM_ECCD_04}
\end{figure}

\begin{figure}
    \centering
    \includegraphics[scale=.9]{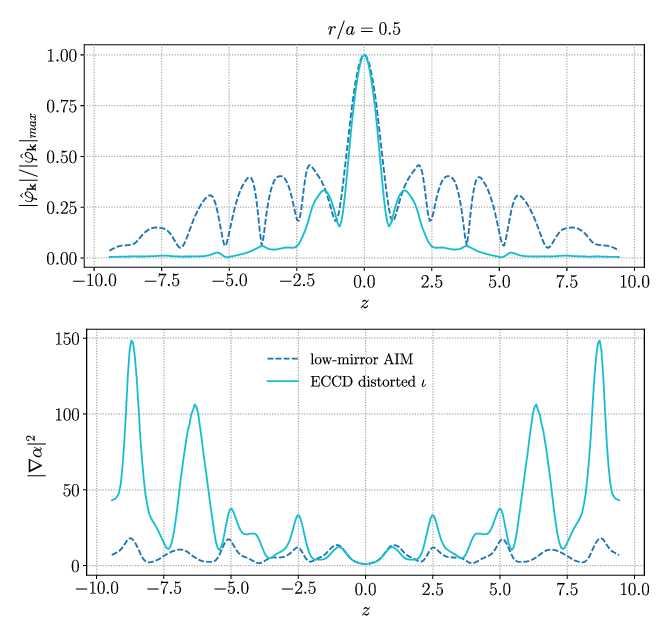}
    \caption{Comparison of the electrostatic potential and $|\nabla\alpha|^2$ structure along the magnetic field line for the low-mirror equilibrium  -- dashed line -- and the ECCD distorted $\iota$ one -- solid line -- at $r/a=0.5$.}
    \label{fig:AIM_ECCD_05}
\end{figure}

In Fig.~\ref{fig:foot_bumpyiota} we show the normalised growth rate and frequency for the most unstable mode as a function of the ion temperature gradient and radial position, for both equilibria. Note that the different points in these plots refer to different radii, with different values of the temperature gradient and magnetic shear at each point. A steepening of the growth rate curve is visible for the ECCD distorted equilibrium at radii where the magnetic shear $\hat{s}_{\text{ECCD}}$ is the largest ($0.45<r/a< 0.6$). This behaviour is favourable as it could produce a higher instability threshold. At smaller radii ($r/a\approx 0.4$), $\hat{s}_{\text{ECCD}}$ decreases and approaches values similar to those in the current-free configuration. The computed growth rates are also similar in the two configurations at these radial positions. We note that a transition to background Floquet modes is observed in both equilibria for radii at which $a/L_{Ti} < 1$. This is evidenced by a jump in the value of the real frequency and by the extended structure of the potential along the field line (see Fig.~\ref{fig:AIM_ECCD_04}).

It thus appears that a large enough global magnetic shear can lead to an increase of the linear instability threshold and, in this sense, to stabilisation. For illustration, we study the structure of the electrostatic potential and flux-tube geometric coefficients along the field line at two different radial positions. The first refers to the location $r/a=0.4$, where the magnitude of the shear is similar for both equilibria: $\hat{s}_{\text{AIM}}=-0.024$ and $\hat{s}_{\text{ECCD}}=-0.019$; and the second to $r/a=0.5$, where the difference between the two equilibria is substantial: $\hat{s}_{\text{AIM}}=-0.057$ and $\hat{s}_{\text{ECCD}}=0.386$. The eigenfunctions are displayed in Fig.~\ref{fig:AIM_ECCD_04} and Fig.~\ref{fig:AIM_ECCD_05}. At the position where the shear is similar ($r/a=0.4$), the two eigenfunctions have similar shapes and extended Floquet modes are observed (Fig.~\ref{fig:AIM_ECCD_04}). In contrast, where the ECCD-distorted $\iota$ equilibrium features a larger and positive shear ($r/a=0.5$) the eigenfunction decays much faster, resulting in stabilisation. 

\begin{figure}
    \centering
    \includegraphics[scale=0.45]{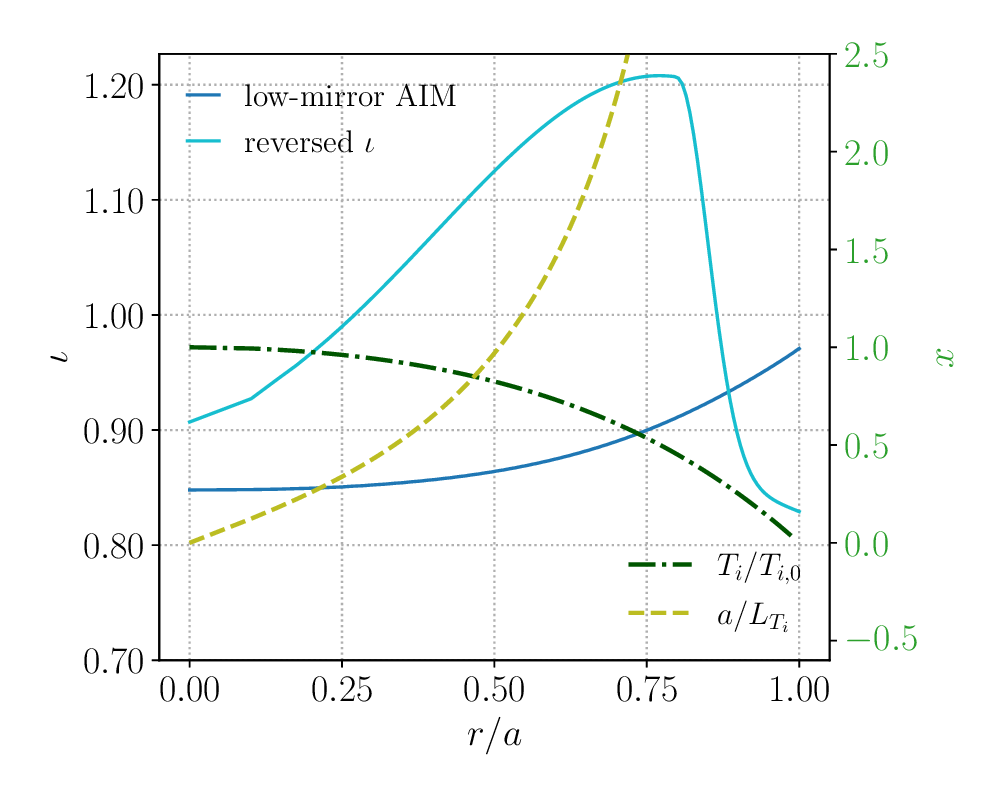}
    \caption{Reversed $\iota$ and low-mirror $\iota$ profiles -- in blue -- with the normalised temperature and ion temperature gradient profiles -- in green.}
    \label{fig:iota_profiles_2}
\end{figure}

The main cause of the eigenfunction localisation can be found in the change of the parallel structure of 
    $$|\nabla\alpha|^2 = |\nabla\theta - \iota \nabla \varphi|^2 - 2 \varphi \iota' \nabla r \cdot (\nabla \theta - \iota \nabla \varphi) + \varphi^2 \iota'^2 |\nabla r|^2  $$ 
with increased shear. This function enters the gyrokinetic equation in the argument of the Bessel function $J_0(k_\perp \rho_i) = J_0(k_\alpha |\nabla \alpha| \rho_i)$. For non-zero shear, $\iota'(r) \ne 0$, it exhibits secular growth along the flux-tube, which is dominated by the last term on the right. Indeed, at $r/a=0.5$, $|\nabla\alpha|^2$ increases roughly quadratically for large $z$, and additionally features spikes that serve to localise the eigenfunction due to FLR suppression \citep{Waltz-Boozer,helander2012stellarator}. This effect has already been observed in tokamaks and in W7-X for ad-hoc constructed $|\nabla\alpha|^2$ profiles \citep{roberg2021calculating}. At $r/a=0.4$, where the global magnetic shear is much smaller, the value of $|\nabla\alpha|^2$ is of comparable amplitude in the two equilibria. 

To further assess the effect of magnetic shear on the ITG growth rate, we prescribe a different equilibrium, where the $\iota$ profile is tailored to possess strongly positive shear in the plasma region where the normalised ion temperature gradient is the largest -- Fig.~\ref{fig:iota_profiles_2}. We will refer to this $\iota$ profile as ‘‘reversed’’.

\begin{figure}
    \centering
    \includegraphics[scale=0.45]{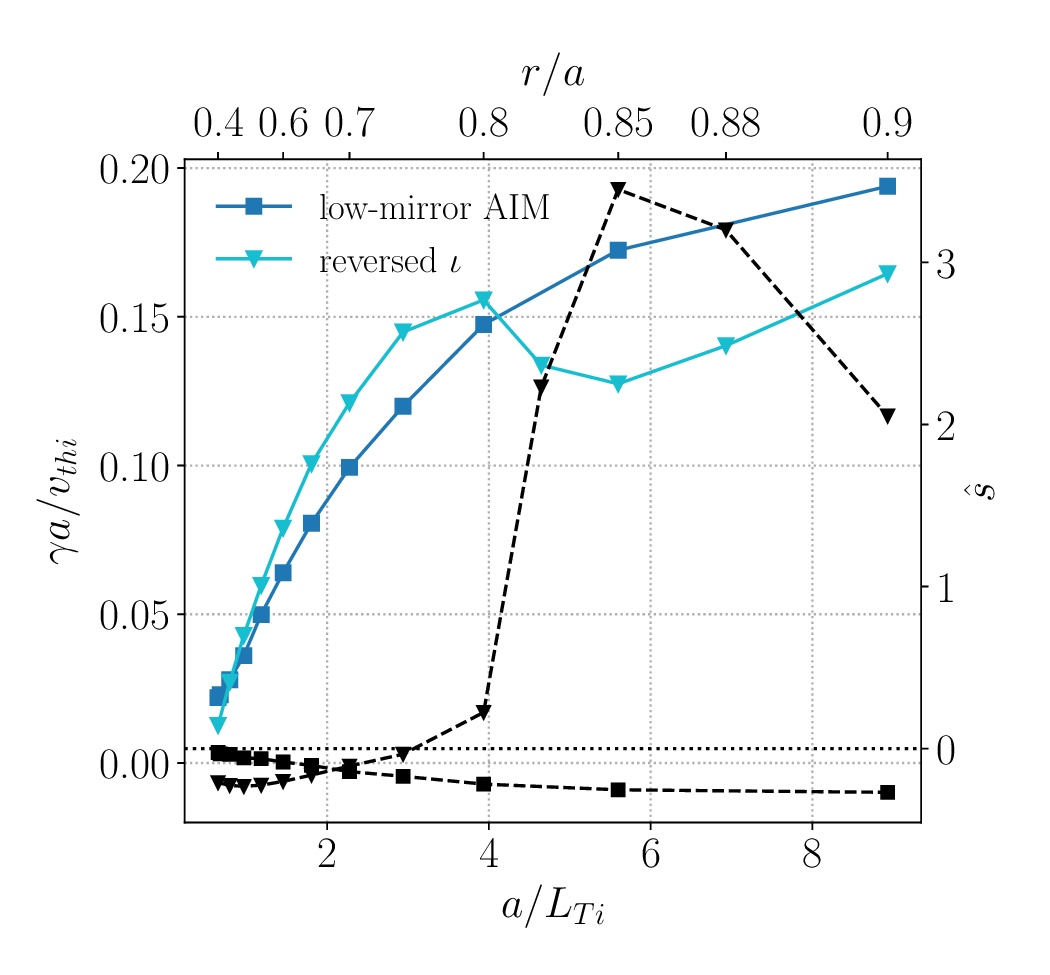}
    \caption{Normalised linear growth rates -- in blue -- as a function of the ion temperature gradient and radial position for the low-mirror configuration and reversed $\iota$ equilibrium. In dashed black the profiles of the magnetic shear for both equilibria.}
    \label{fig:foot_reviota}
\end{figure}

We first run simulations of plasmas with an ion temperature gradient only, and compare them with similar simulations run in the ordinary low-mirror configuration (Fig.~\ref{fig:foot_reviota}). We follow the same procedure as above: we run each simulation at a different radial position and change the temperature gradient and $\tau$ consistently with the plasma profiles. A summary of all the radial locations, shear, gradients and $\tau$ values, together with the resulting growth rates is given in table \ref{tab:simulations_2} in Appendix \ref{sec:appendix}. In the reversed $\iota$ equilibrium, the growth rate shows a non-monotonic trend with increasing ion temperature gradient (and $r/a$). The reduction is up to 20\% when the magnetic shear $\hat{s}_{\text{rev}}$ reaches its maximum value at $r/a=0.85$. Apparently, the stabilising effect of magnetic shear more than offsets the destabilising effect of a larger temperature gradient. Another interesting feature is the steepening of the curve at lower gradients. Compared to the result in the low-mirror configuration, the reversed $\iota$ one seems to show a larger linear threshold and no presence of Floquet background modes. 

As before, we look at the structure of the electrostatic potential and the geometric coefficients along the magnetic field line. We study the radial position $r/a=0.85$, where the growth rate shows the maximum reduction. At this position the values of the magnetic shear in the two equilibria are, respectively, $\hat{s}_{\text{AIM}}=-0.254$ and $\hat{s}_{\text{rev}}=3.450$. We then consider a case where Floquet modes are observed in the low-mirror equilibrium but not in the reversed $\iota$ one, namely $r/a=0.4$. The values of the magnetic shear are here: $\hat{s}_{\text{AIM}}=-0.024$ and $\hat{s}_{\text{rev}}=-0.210$. 

\begin{figure}
    \centering
    \includegraphics[scale=1.5]{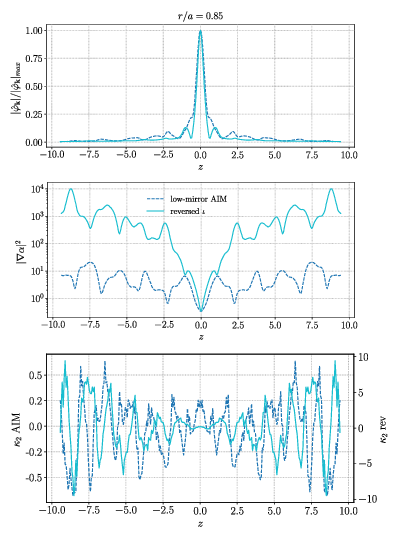}
    \caption{Comparison of the electrostatic potential, the quantity $|\nabla\alpha|^2$ on a semi-logarithmic scale, and $\kappa_2$ structure along the magnetic field line for the low-mirror equilibrium  -- dashed line -- and the reversed $\iota$ one  -- solid line -- at $r/a=0.85$.}
    \label{fig:AIM_rev_085}
\end{figure}

\begin{figure}
    \centering
    \includegraphics[scale=1.5]{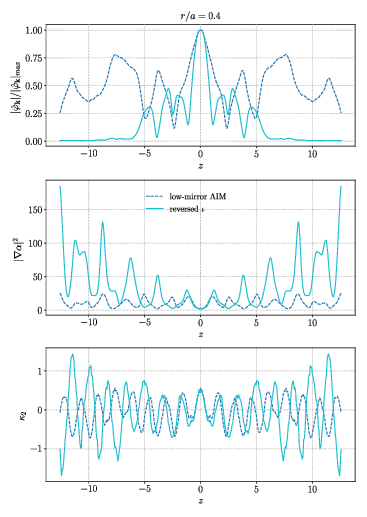}
    \caption{Comparison of the electrostatic potential, the quantity $|\nabla\alpha|^2$, and $\kappa_2$ structure along the magnetic field line for the low-mirror equilibrium  -- dashed line -- and the reversed $\iota$ one -- solid line -- at $r/a=0.4$.}
    \label{fig:AIM_rev_4}
\end{figure}

In the first case ($r/a=0.85$), the eigenfunctions are very localised for both equilibria (Fig.~\ref{fig:AIM_rev_085}). The gradient is large and the mode is fluid-like. When looking at the structure of $|\nabla\alpha|^2$, however, we note a big difference between the two equilibria. The reversed $\iota$ one features peaks that reach amplitudes two orders of magnitude higher than the low-mirror one. In this case, $|\nabla\alpha|^2$ plays no role in the localisation of the mode but rather in its stabilisation, possibly still through FLR suppression. Similarly, the magnetic drift $\kappa_2$ exhibits differences both in magnitude and in location. This could also play a role both in the localisation and stabilisation of the mode. Note the two different scales in the plots for the two equilibria.

In the second case ($r/a=0.4$), the low-mirror equilibrium features a Floquet mode, extended along the field line, while the reversed $\iota$ one shows a more localised mode (Fig.~\ref{fig:AIM_rev_4}). This explains the absence of a foot in the growth rate for the latter case. When looking at the structure of $|\nabla\alpha|^2$, we notice that we can again ascribe the localisation of the eigenfunction to an increase in the magnitude of $|\nabla\alpha|^2$ at the extremes of the flux-tube. It is important to notice that in this case the sign of the shear is negative, but it is nevertheless one order of magnitude larger in magnitude than the one of the low-mirror configuration. We thus conclude that it is the magnetic shear magnitude that plays a role in the increase of the linear instability threshold, not its sign. Finally, the structure of the magnetic drift $\kappa_2$ also shows differences, even if less pronounced than in the case previously studied. 

\begin{figure}
    \centering
    \includegraphics[scale=0.68]{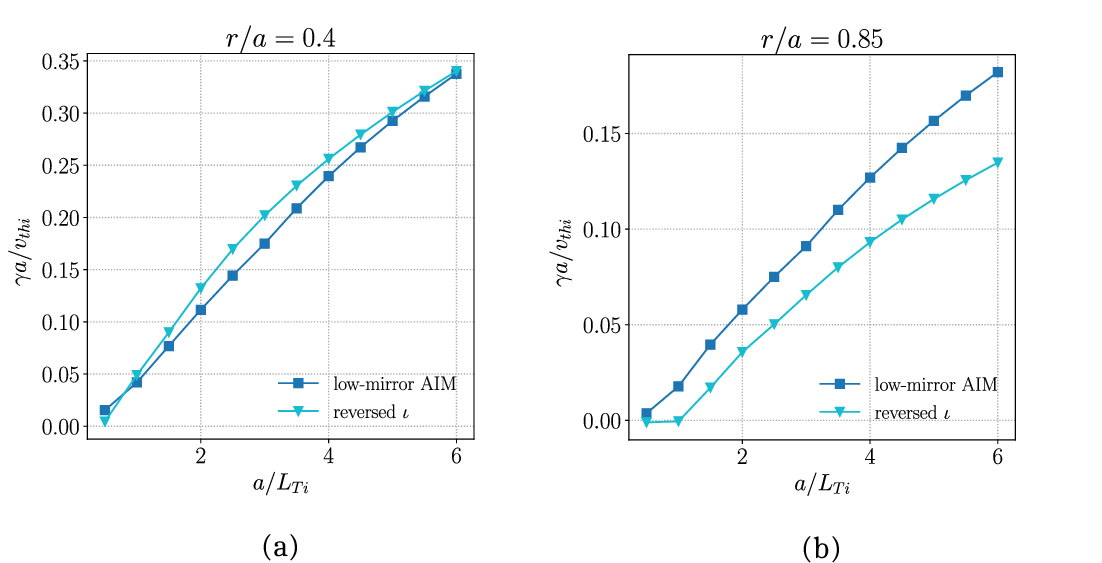}
    \caption{Normalised linear growth rates for the low-mirror configuration and reversed $\iota$ equilibrium at (a) $r/a=0.4$ and (b) $r/a=0.85$. Featuring a near-marginality growth rate steepening in (a) and a total suppression of Floquet modes in (b).}
    \label{fig:fixed}
\end{figure}

An inconsistent $a/L_{Ti}$ scan performed at the just discussed radial positions confirms the predominant role of the shear in the modification of the ITG growth rate in Fig.~\ref{fig:foot_reviota}. At $r/a=0.4$ (Fig.~\ref{fig:fixed}a) in the reversed $\iota$ equilibrium, we observe a near-threshold steepening of the growth rate curve due to Floquet modes stabilisation (Fig.~\ref{fig:AIM_rev_4}). At $r/a=0.85$ (Fig.~\ref{fig:fixed}b) both the toroidal ITG branch and the Floquet ones are stabilised. In particular, the latter is completely damped, yielding negative growth rate values for $a/L_{Ti}<1.5$.

We conclude that increasing the absolute value of the magnetic shear could be a way to affect both the far and near threshold behaviour of ITG modes. In particular, it seems to be an effective way to suppress the Floquet modes, otherwise always present, and thus to increase the linear instability threshold in low-shear devices. 

We note that the favourable instability threshold increase due to steepening of the growth rate trend can come with the price of destabilisation at other radial positions. This happens where the distorted $\iota$ equilibria feature similar or smaller magnetic shear magnitude to the low-mirror configuration (see both Fig.~\ref{fig:foot_bumpyiota} and Fig.~\ref{fig:foot_reviota}). In the following nonlinear analysis, we tackle the effect this destabilisation can have on transport. 

In order to understand the effect of a distorted iota profile on transport, we run nonlinear simulations for the AIM and reversed $\iota$ equilibria. We take two representative cases: one where the magnitude of the shear in the reversed $\iota$ equilibrium is the largest ($r/a=0.85$ with $\hat{s}_{\text{rev}}=3.450$) and one where it is the smallest ($r/a=0.75$ with $\hat{s}_{\text{rev}}=-0.035$). In the latter, in particular, the reversed $\iota$ equilibrium exhibits a larger linear growth rate than in the low-mirror configuration and we are interested in observing whether this destabilisation is reflected also nonlinearly. For the same radial positions, the low-mirror configuration has $\hat{s}_{\text{AIM}}=-0.171$ for $r/a=0.75$ and $\hat{s}_{\text{AIM}}=-0.254$ for $r/a=0.85$. In the two radial positions, $a/L_{Ti}=2.49$ and $a/L_{Ti}=5.60$ for, respectively, $r/a=0.75$ and $r/a=0.85$. We expect to observe fluid-like turbulence in both cases and for this reason we choose to run the simulations with the resolutions reported in the first row of table \ref{tab:resolutions}. We set $a/L_n=a/L_{T_e}=0$ and $\tau$ consistent with plasma profiles, as reported in table \ref{tab:simulations_2} in Appendix~\ref{sec:appendix}. 

We report the average saturated ion energy flux in table \ref{tab:nonlinear_shear}. At the radius where the reversed $\iota$ equilibrium features a large, positive shear ($r/a=0.85$), the ion energy flux is about 8 times smaller than in the low-mirror configuration. The result confirms the linearly observed reduction. The linear trend is also confirmed at $r/a=0.75$, where a larger growth rate in the reversed $\iota$ configuration reflects in a larger ion energy flux. The difference with the low-mirror configuration is, however, less than a 1.5 factor. The desired maximisation of the stabilising shear effect where the gradients are the steepest is thus confirmed also nonlinearly. It is plausible to conjecture that the mild linear destabilisation at around $r/a=0.75$ is not as important as the larger stabilising effect just discussed. 

\begin{table}
\begin{center}
\def~{\hphantom{0}}
\begin{tabular}{p{0.12\textwidth}>{\centering}p{0.12\textwidth}>{\centering\arraybackslash}p{0.12\textwidth}}
	$\mathbf{r/a}$ & \textbf{AIM} & \textbf{reversed} $\mathbf{\iota}$ \\
	\hline
	0.75 & 4.86 & 6.71\\
	0.85 & 15.44 & 1.92\\
\end{tabular}
\caption{Resulting average ion energy flux $Q_i/Q_{gB}$ from nonlinear simulations in the low-mirror (AIM) and reversed $\iota$ configurations at two different radial locations.}
\label{tab:nonlinear_shear}
\end{center}
\end{table}

\begin{figure}
    \centering
    \includegraphics[scale=0.42]{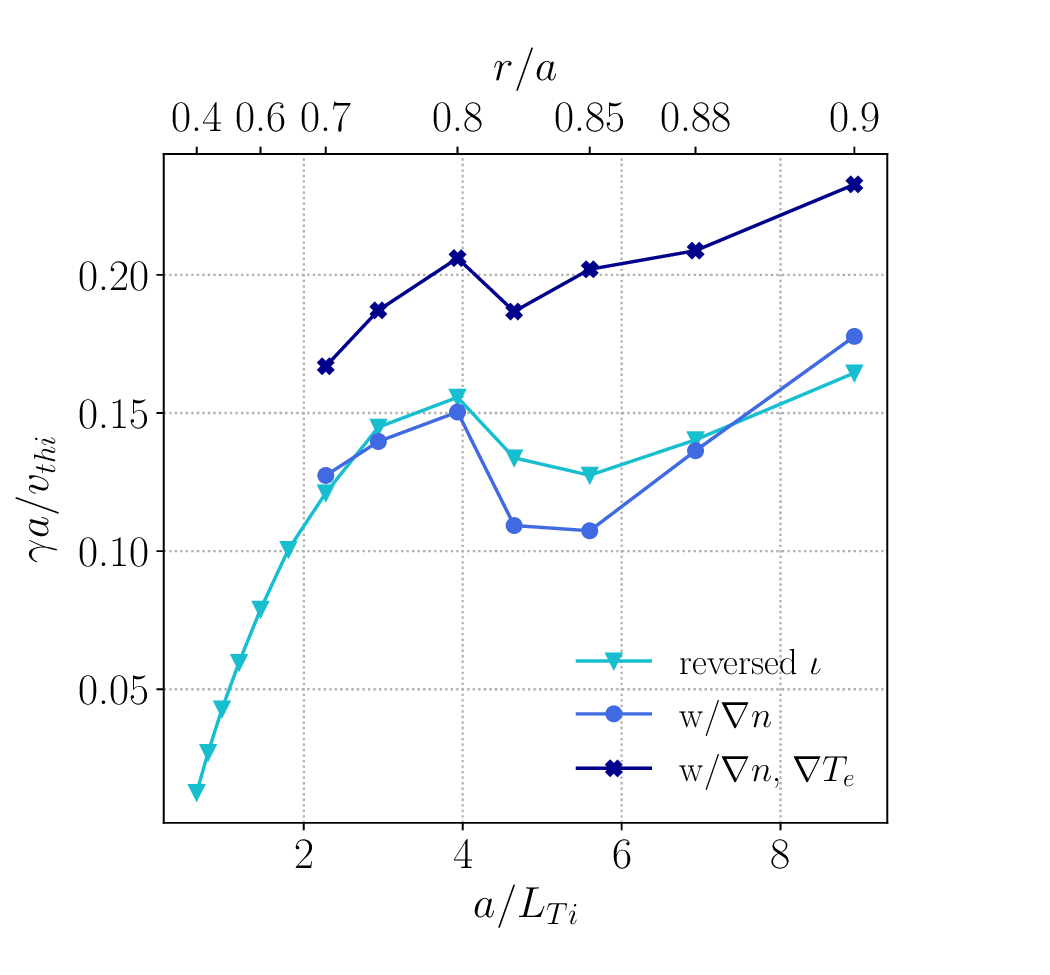}
    \caption{Normalised linear growth rates -- in blue -- as a function of the ion temperature gradient and radial position for the reversed $\iota$ equilibrium with the effects of a finite density and electron temperature gradient.}
    \label{fig:foot_reviota_2}
\end{figure}

Finally, we take into account the effects of finite density and electron temperature gradients (Fig.~\ref{fig:foot_reviota_2}). We switch them on one by one and re-run the linear
simulations at the radial locations where the shear effect is the strongest. All the gradients are changed at every radial location, consistently with plasma profiles. A summary of all the radial locations, shear, gradients and $\tau$ values, together with the resulting growth rates is given in table \ref{tab:simulations_3} in Appendix \ref{sec:appendix}. As noted in the diagrams in the first section and in previous works \citep{helander2015advances,Alcuson_2020}, the addition of a non-zero density gradient in W7-X leads to a reduction in the growth rate at the radial locations where the shear is the strongest. If a finite electron temperature gradient is also added, the growth rate curve is shifted upwards, i.e., the plasma becomes more unstable. A non-monotonic trend of the growth rate is, however, still observed thanks to the stabilising action of a large shear. It is worth mentioning that only ion length scales are included in these simulations and the growth rates reported in Fig.~\ref{fig:foot_reviota_2} all refer to fluid-like ITG modes. We conclude that, also in the most realistic case, the tweaking of the $\iota$ profile seems to reduce the ITG instability, although in a less pronounced fashion.   

\section{Discussion and conclusions}
\label{sec:conclusions}

We have characterised electrostatic gyrokinetic instabilities and turbulence in W7-X, with particular attention given to their character close to the linear stability threshold. The presence and character of ion temperature and density gradient driven instabilities can conveniently be summarised in stability diagrams. In contrast to a previous work \citep{Alcuson_2020}, we have constructed different diagrams showing instabilities with linear growth rates at different spatial scales and rotating in different directions. This method ensures that no ion temperature or density gradient driven instability is overlooked, especially when it is not the most unstable but it is destabilised at scales that are more relevant for transport. An important example is the observation of universal modes \citep{Coppi_1977,smoliakovkishimoto,landreman2015,helander-plunk2015} for large enough density gradients. These modes were previously observed numerically in W7-X simulations \citep{costello2023universal}, but were thought to be overwhelmed by faster growing instabilities in the presence of an ion temperature gradient. Through the stability diagrams we have here shown how these modes are stabilised by the presence of an ion temperature gradient but are always present, despite being slightly more stable than ITGs or iTEMs. More importantly, they are characterised by wavelengths longer than the ion Larmor radius and a quasi-linear estimate of the turbulent heat flux suggests they are relevant for transport in W7-X.   

We have paid particular attention to the geometric nature of ITG instabilities close to the linear stability threshold. For small ion temperature gradients, the dominant ITG modes are of the Floquet-type \citep{bhatt,candywaltzrosen,jackbryanfloq,zocco2018threshold} and produce small but non-zero energy transport \citep{PhysRevE.106.L013202}. These modes are greatly extended along field lines, particularly close to the linear instability threshold. Long flux-tubes are thus necessary in numerical simulations to resolve the slowly decaying mode structure along the magnetic field. Nonlinearly, these modes behave differently from fluid-like ones. In particular, they are more radially localised, which may explain why zonal flows are not efficient in suppressing Floquet-type turbulence \citep{PhysRevE.106.L013202}. A range of diagnostics and numerical experiments corroborate this hypothesis. We have also observed a general steepening of fluctuation spectra with decreasing ion temperature gradient, which might be the sign of a transition to the sub-$\rho_i$ gyrokinetic turbulent regime predicted by \citep{Schekochihin_2008}.

There is good reason to believe that extended, Floquet-type modes are ubiquitous for small magnetic shear. Indeed, instabilities that extend far along the magnetic field have been observed in tokamak simulations too, both with adiabatic \citep{zocco2018threshold} and kinetic electrons \citep{volvcokas2022ultra}. We have observed them in near-marginal ITG numerical simulations with kinetic electrons performed in different W7-X equilibria: the high-mirror (KJM) and low-mirror (AIM) magnetic configurations, but also in modified W7-X magnetic equilibria. 

W7-X is inherently a low-shear device and thus prone to the formation of such elongated modes, but the global magnetic shear can be locally enhanced by electron cyclotron current drive (ECCD) \citep{zanini2020eccd, zanini2021confinement, aleynikova2021model, Zocco_2021}. The effect of the driven toroidal current is a distortion of the $\iota$ profile, which can be modelled by tailoring the $\iota$-profile of the magnetic equilibrium whilst keeping the density and temperature profiles fixed. The presence of large magnetic shear over a carefully chosen plasma region causes the growth rate to depend non-monotonically on radius, even in the most realistic scenarios where both the finite density and electron temperature gradients are taken into account. The beneficial effect is not only linear. An equilibrium with a strong, positive shear can show locally ion energy fluxes eight times smaller than an equilibrium with a non-distorted $\iota$-profile. In such tailored magnetic equilibria, the dependence of the instability growth rate on the temperature gradient suggests an increase in the linear stability threshold and, indeed, extended modes can be successfully suppressed. Manipulation of the $\iota$ profile can thus be exploited as a tool for ITG stabilisation in W7-X. In particular, the small but finite level of transport caused by Floquet-type turbulence could be reduced and the ITG critical gradient increased. 

\begin{acknowledgements}
This work has been carried out within the framework of the EUROfusion Consortium, funded by the European Union via the Euratom Research and Training Programme
(Grant Agreement N. 101052200 — EUROfusion). Views and opinions expressed are however those of the author(s) only and do not necessarily reflect those of the
European Union or the European Commission. Neither the European Union nor the European Commission can be held responsible for them. The \texttt{stella} simulations have been performed on the HPCs Draco, Raven (Germany) and Marconi (Italy). The authors report no conflict of interest.
\end{acknowledgements}

\appendix
\section{}
\label{sec:appendix}

\begin{table}
\begin{center}
\def~{\hphantom{0}}
\begin{tabular}{p{0.12\textwidth}>{\centering}p{0.12\textwidth}>{\centering}p{0.12\textwidth}>{\centering}p{0.12\textwidth}>{\centering}p{0.12\textwidth}>{\centering}p{0.12\textwidth}>{\centering\arraybackslash}p{0.12\textwidth}}
	$\mathbf{r/a}$ & $\mathbf{\hat{s}_{\text{ECCD}}}$ & $\mathbf{\hat{s}_{\text{AIM}}}$ & $\mathbf{\tau}$ & $\mathbf{a/L_{Ti}}$ & $\mathbf{\hat{\gamma}_{\text{ECCD}}}$ & $\mathbf{\hat{\gamma}_{\text{AIM}}}$\\
	\hline
	0.40 & -0.019 & -0.024 & 0.46 & 0.65 & 0.021 & 0.022\\
	0.41 & 0.029 & -0.033 & 0.47 & 0.68  & 0.022 & 0.023\\
    0.45 & 0.223 & -0.035 & 0.48 & 0.80 & 0.021 & 0.028\\
    0.50 & 0.386 & -0.057 & 0.50 & 0.97  & 0.031 & 0.036\\
    0.55 & 0.386 & -0.062 & 0.53 & 1.18  & 0.053 & 0.050\\
    0.60 & 0.253 & -0.083 & 0.56 & 1.46  & 0.073 & 0.064\\
    0.65 & 0.084 & -0.104 & 0.59 & 1.81  & 0.090 &  0.081\\
    0.70 & -0.047 & -0.142 & 0.63 & 2.28  & 0.105 & 0.100\\
\end{tabular}
\caption{Set of physical parameters for the linear simulations performed in the ECCD distorted equilibrium and the low-mirror (AIM) equilibrium.}
\label{tab:simulations}
\end{center}
\end{table}

\begin{table}
\begin{center}
\def~{\hphantom{0}}
\begin{tabular}{p{0.12\textwidth}>{\centering}p{0.12\textwidth}>{\centering}p{0.12\textwidth}>{\centering}p{0.12\textwidth}>{\centering}p{0.12\textwidth}>{\centering}p{0.12\textwidth}>{\centering\arraybackslash}p{0.12\textwidth}}
	$\mathbf{r/a}$ & $\mathbf{\hat{s}_{\text{rev}}}$ & $\mathbf{\hat{s}_{\text{AIM}}}$ & $\mathbf{\tau}$ & $\mathbf{a/L_{Ti}}$ & $\mathbf{\hat{\gamma}_{\text{rev}}}$ & $\mathbf{\hat{\gamma}_{\text{AIM}}}$ \\
	\hline
	0.40 & -0.210 & -0.024 & 0.46 & 0.65 & 0.013 & 0.022 \\
    0.45 & -0.227 & -0.035 & 0.48 & 0.80 & 0.027 & 0.028 \\
    0.50 & -0.233 & -0.057 & 0.50 & 0.97 & 0.043 & 0.036 \\
    0.55 & -0.225 & -0.062 & 0.53 & 1.18 & 0.060 & 0.050 \\
    0.60 & -0.203 & -0.083 & 0.56 & 1.46 & 0.079 & 0.064 \\
    0.65 & -0.163 & -0.104 & 0.59 & 1.81 & 0.101 & 0.081 \\
    0.70 & -0.108 & -0.142 & 0.63 & 2.28 & 0.121 & 0.099 \\
    0.75 & -0.035 & -0.171 & 0.68 & 2.94 & 0.145 & 0.120 \\
    0.80 & 0.224 & -0.219 & 0.74 & 3.93 & 0.156 &  0.147 \\
    0.825 & 2.231 &      & 0.77 & 4.65 & 0.134 & \\
    0.85 & 3.450 & -0.254 & 0.81 & 5.60 & 0.127 & 0.172 \\
    0.875 & 3.200 &      & 0.84 & 6.93 & 0.140 & \\
    0.90 & 2.052 & -0.270 & 0.89 & 8.93 & 0.165 & 0.194 \\
\end{tabular}
\caption{Set of physical parameters for the linear simulations performed in the reversed $\iota$ equilibrium and the low-mirror (AIM) equilibrium with $a/L_{T_i}$ only.}
\label{tab:simulations_2}
\end{center}
\end{table}

\begin{table}
\begin{center}
\def~{\hphantom{0}}
\begin{tabular}{p{0.10\textwidth}>{\centering}p{0.10\textwidth}>{\centering}p{0.10\textwidth}>{\centering}p{0.10\textwidth}>{\centering}p{0.10\textwidth}>{\centering}p{0.10\textwidth}>{\centering}p{0.10\textwidth}>{\centering\arraybackslash}p{0.10\textwidth}}
	$\mathbf{r/a}$ & $\mathbf{\hat{s}_{\text{rev}}}$ & $\mathbf{\tau}$ & $\mathbf{a/L_{Ti}}$ & $\mathbf{a/L_n}$ & $\mathbf{\hat{\gamma}_{\nabla n}}$ & $\mathbf{a/L_{Te}}$ & $\mathbf{\hat{\gamma}_{\nabla Te}}$\\
	\hline
    0.70 & -0.108 & 0.63 & 2.28 & 1.15 & 0.127 & 3.68 & 0.167 \\
    0.75 & -0.035 & 0.68 & 2.94 & 1.52 & 0.140 & 4.47 & 0.187 \\
    0.80 & 0.224 & 0.74 & 3.93 & 2.01 & 0.150 & 5.62 & 0.206 \\
    0.825 & 2.231 & 0.77 & 4.65 & 2.33 & 0.109 & 6.42 & 0.187 \\
    0.85 & 3.450 & 0.81 & 5.60 & 2.71 & 0.107 & 7.46 & 0.202 \\
    0.875 & 3.200 & 0.84 & 6.93 & 3.18 & 0.136 & 8.88 & 0.209 \\
    0.90 & 2.052 & 0.89 & 8.93 & 3.77 & 0.178 & 11.00 & 0.233 \\
\end{tabular}
\caption{Set of physical parameters for the linear simulations performed in the reversed $\iota$ equilibrium with $a/L_{Ti}$, $a/L_{n}$ and $a/L_{Te}$.}
\label{tab:simulations_3}
\end{center}
\end{table}

\clearpage
\bibliographystyle{jpp}
\bibliography{main}

\end{document}